\documentclass[prd,twocolumn,nofootinbib]{revtex4}
\usepackage{graphicx}
\usepackage{latexsym}

\begin{document}

\title{The High Redshift Integrated Sachs-Wolfe Effect}

\author{Jun-Qing Xia${}^1$}
\email{xia@sissa.it}
\author{Matteo Viel${}^{2,3}$}
\email{viel@oats.inaf.it}
\author{Carlo Baccigalupi${}^{1,3}$}
\email{bacci@sissa.it}
\author{Sabino Matarrese${}^{4,5}$}
\email{sabino.matarrese@pd.infn.it}

\affiliation{${}^1$Scuola Internazionale Superiore di Studi
Avanzati, Via Beirut 2-4, I-34014 Trieste, Italy}

\affiliation{${}^2$INAF-Osservatorio Astronomico di Trieste, Via
G.B. Tiepolo 11, I-34131 Trieste, Italy}

\affiliation{${}^3$INFN/National Institute for Nuclear Physics, Via
Valerio 2, I-34127 Trieste, Italy}

\affiliation{${}^4$Dipartimento di Fisica ``G. Galilei'',
Universit\`a di Padova, Via Marzolo 8, I-35131 Padova, Italy}

\affiliation{${}^5$INFN, Sezione di Padova, Via Marzolo 8, I-35131
Padova, Italy}


\begin{abstract}

In this paper we rely on the quasar (QSO) catalog of the Sloan
Digital Sky Survey Data Release Six (SDSS DR6) of about one million
photometrically selected QSOs to compute the Integrated Sachs-Wolfe
(ISW) effect at high redshift, aiming at constraining the behavior
of the expansion rate and thus the behaviour of dark energy at those
epochs. This unique sample significantly extends previous catalogs
to higher redshifts while retaining high efficiency in the selection
algorithm. We compute the auto-correlation function (ACF) of QSO
number density from which we extract the bias and the stellar
contamination. We then calculate the cross-correlation function
(CCF) between QSO number density and Cosmic Microwave Background
(CMB) temperature fluctuations in different subsamples: at high
$z>1.5$ and low $z<1.5$ redshifts and for two different choices of
QSO in a conservative and in a more speculative analysis. We find an
overall evidence for a cross-correlation different from zero at the
$2.7\,\sigma$ level, while this evidence drops to $1.5\,\sigma$ at
$z>1.5$. We focus on the capabilities of the ISW to constrain the
behaviour of the dark energy component at high redshift both in the
$\Lambda$CDM and Early Dark Energy cosmologies, when the dark energy
is substantially unconstrained by observations. At present, the
inclusion of the ISW data results in a poor improvement compared to
the obtained constraints from other cosmological datasets. We study
the capabilities of future high-redshift QSO survey and find that
the ISW signal can improve the constraints on the most important
cosmological parameters derived from Planck CMB data, including the
high redshift dark energy abundance, by a factor $\sim 1.5$.

\end{abstract}


\maketitle

\section{Introduction}
\label{int}

The cosmic microwave background (CMB) measurements of temperature
anisotropies and polarization \cite{Komatsu:2008hk} and the
redshift-distance measurements of Type Ia Supernovae (SNIa) at $z<2$
\cite{Kowalski:2008ez}, have established that the universe is
undergoing an accelerated phase of expansion and that its total
energy budget is dominated by a dark energy component.  The nature
of this component is still unknown and many observational probes
have been proposed to test its properties and redshift evolution
either in the standard $\Lambda$-Cold Dark Matter ($\Lambda$CDM),
modified gravity or quintessence models (for a review see Ref.
~\cite{copeland}). While the CMB is a powerful cosmological probe of
the universe at $z\sim 1100$, the anisotropies present in CMB data
(kinetic and thermal Sunyaev-Zeldovich effects, weak lensing and the
Integrated Sachs-Wolfe effect) contain precious information on the
large scale structure that formed at much lower redshift. These
effects can be studied and detected by cross-correlating CMB data
with tracers of the large scale structure (LSS) such as galaxies or
quasars. Here, we will focus on the Integrated Sachs-Wolfe effect
\cite{sachswolfe67}.

The first investigations of CMB-LSS cross-correlations were made in
Refs.~\cite{crittendenturok96,boughn98} using X-ray observations
(HEAO data) and CMB data from the COBE satellite that allowed to
constrain the bias of X-ray sources and put upper limits on the
amount of a cosmological constant energy density. In a more recent
series of works, a similar set of analyses have been carried which
relied on CMB data from the WMAP satellite and a variety of LSS
probes such as NVSS (NRAO VLA Sky Survey) radio galaxies
\cite{noltaetal04,raccanellietal08}, the Two Micron All Sky Survey
(2MASS) \cite{afshordietal04,rassat07}, galaxies from Sloan Digital
Sky Survey (SDSS) DR4 and DR5 \cite{cabre06,cabre07}.  There is
overall agreement between the different groups in finding an
evidence for a positive ISW signal at the $\sim 3\,\sigma$
confidence level.  Even more recently, several groups have started
to analyze the CMB-LSS correlations in a more comprehensive
framework fully exploiting the capabilities of the SDSS (DR5, DR6),
addressing the many possible systematic effects involved, modeling
the statistical error bars and covariance properties of the data
with different methods and combining the different tracers in the
using Monte Carlo Markov Chains estimators for deriving the
cosmological parameters (e.g.
\cite{gianna06,gianna08,hoetal08,hirataetal08}).

Although the results obtained from the CMB-LSS are promising and
have provided independent evidences for the presence of a dark
energy component, their quantitative use as cosmological probes is
still to be fully exploited. The timely convergence of future LSS
surveys like SDSS-III \cite{boss07,bigboss09} and high resolution
CMB experiment like Planck\footnote{Available at
http://www.rssd.esa.int/planck/.} will offer the opportunity to
further explore the valuable information of the ISW  and in general
of several other LSS-CMB cross-correlations in a quantitative way
(e.g. \cite{vallinottoetal09}). Present efforts are particularly
concentrated on quantifying several systematic effects that are
possibly affecting the ISW signal such as noise by local variance
\cite{frommertetal08}, redshift space distortions \cite{rassat09},
non-linear contributions \cite{caietal09}, uncertainties in the bias
estimates \cite{schaefer09} and contributions due to voids and
clusters to the overall ISW signal \cite{granett08}.

However, once the systematic effects will be under control the ISW
is likely to become a powerful cosmological probe able to constrain
for example the contribution of massive neutrinos \cite{lesg08} or
evolution of the dark energy component at high redshift (e.g.
~\cite{pogosian06}), which is poorly constrained by observations at
the present \cite{seljak_etal_2005}. In this paper we focus on this
latter issue and push the ISW capabilities to the highest possible
redshift regime in a way similar to that used by
Ref.~\cite{gianna06}, using the recently released SDSS DR6-QSO
catalog of about one million photometrically selected QSOs in order
to give constraints on some dark energy models.

Among all the dynamical dark energy models, we will consider early
dark energy (EDE) ones, in which a small fraction of dark energy is
present up to the last scattering surface (lss), unlike $\Lambda$CDM
for which $\Omega_{\rm DE}(z_{\rm lss})\simeq0$, The differences
between early dark energy models and pure $\Lambda$CDM are
particularly evident at high redshifts, over a large fraction of the
cosmic time, when the first structures form.  EDE has been shown to
influence the growth of cosmic structures (both in the linear and in
the non-linear regime), to change the age of the universe, to have
an influence on CMB physics, to impact on the reionization history
of the universe, to modify the statistics of giant arcs in strong
cluster lensing statistics (e.g.
\cite{doran01,linderjenkins03,dolag04,mainini04,bartelmann06,fedeli07,
acquaviva_baccigalupi_2006,smith_etal_2006,crociani08,grossi08,francis08a,francis08b,mota08}).
Recently some EDE models have been investigated by
Refs.~\cite{xiaviel09,hollenstein09}, focussing in particular on
using present and future weak lensing observables and measurements
of the growth factors of density perturbations obtained via
Lyman-$\alpha$ forest observations.

The structure of the paper is as follows: in Sec.~\ref{isw} and
Sec.~\ref{qso} we briefly review the theoretical background of the
ISW effect and describe the quasar catalog used, respectively.
Sec.~\ref{acf} and Sec.~\ref{ccf} contain the analysis of ACF of QSO
number density and CCF between QSO number density and CMB
temperature fluctuations. In Sec.~\ref{method} we present the
theoretical framework of the early dark energy model and the
datasets we used. Sec.~\ref{results} contains the bulk of our
results, while Sec.~\ref{futresult} is dedicated to forecasting with
future datasets. We conclude with a discussion in
Sec.~\ref{summary}.

\section{ISW Effect}
\label{isw}

In this section we briefly review the basics of ISW effect focussing
on its cross-correlation with the number density of astrophysical
sources (e.g. Refs.~\cite{afshordietal04,peiris00,cooray02}).

The temperature anisotropy due to the ISW effect is expressed as an
integral of the time derivative of the gravitational potential
$\Phi$ over conformal time $\eta$
\begin{equation}
{\frac{\Delta T}{T}}^{\rm ISW}(\hat{\bf
n})=-2\int{\dot{\Phi}[\eta,\hat{\bf
n}(\eta_0-\eta)]d\eta}~.\label{isweq}
\end{equation}
For scales within the horizon, we can relate the gravitational
potential $\Phi$ to the comoving density field $\delta$ by Poisson
equation:
\begin{eqnarray}
\nabla^2\Phi(x)&=&\frac{4\pi
G\rho_{\rm m}}{a}\delta_{\rm m}(x)~~~~~~\Rightarrow~~\nonumber\\
\Phi({\bf
k},z)&=&-\frac{3H^2_0}{2c^2}\Omega_{\rm m}(1+z)\frac{\delta_{\rm m}({\bf
k},z)}{k^2}~,\label{possion}
\end{eqnarray}
where $\Omega_{\rm m}$ is the ratio of the matter density to the
critical density today, $H_0$ is the Hubble constant today, $c$ is
the speed of light, $z$ is the redshift, and $k$ is the comoving
wave number. From Eqs.~(\ref{isweq},\ref{possion}) one can
appreciate that when a CMB photon falls into a gravitational
potential well, it gains energy, while it loses energy when it
climbs out of a potential well. These effects exactly cancel if the
potential is time independent, such as the matter dominated era
($\delta_{\rm m}\sim{a}$) in which the gravitational potential stays
constant, $\dot\Phi=0$ and no ISW is produced. However, when dark
energy or curvature become important at later times, the potential
evolves as the photon passes through it. In this case,
$\dot\Phi\neq0$ and additional CMB anisotropies will be produced.

The ISW effect of interest here is the one produced at relatively
late time, when the dark energy component is dominating the universe
density budget, causing a change in the time dependence of the
expansion rate, departing from pure matter dominance, and a
consequent time evolution of the gravitational potentials. An early
ISW, not considered here, is injected soon after decoupling, when
the expansion rate time dependence is in the transition between
radiation and matter dominance. Observing the late-time ISW can be a
powerful way of probing dark energy and its evolution. However, the
most significant ISW effect contributes to the CMB anisotropies on
large scales that are strongly affected by the cosmic variance.
Fortunately, this problem can be solved by the cross-correlation
between ISW temperature fluctuation and the density of astrophysical
objects like galaxies or quasars (in the following calculations we
will use the quasar catalog). The observed quasar (QSO) density
contrast in a given direction $\hat{\bf n}_1$ will be:
\begin{eqnarray}
\delta_{\rm q}(\hat{\bf n}_1)&=&\int{f(z)\delta_{\rm m}(\hat{\bf
n}_1,z)dz}\nonumber\\
&=&\int{b_{\rm q}(z)\frac{dN}{dz}(z)\delta_{\rm m}(\hat{\bf
n}_1,z)dz}~,
\end{eqnarray}
where $b_{\rm q}(z)$ is an assumed scale-independent bias factor
relating the quasar overdensity to the mass overdensity,
$\delta_{\rm q}=b_{\rm q}\delta_{\rm m}$, $dN/dz$ is the normalized
selection function of the survey. Since the density $\delta_{\rm m}$
is related to the gravitational potential $\Phi$, the observed
galaxy density will be correlated with the ISW temperature in the
nearby direction $\hat{\bf n}_2$:
\begin{equation}
\frac{\Delta T}{T}(\hat{\bf n}_2)=-2\int{\frac{d\Phi}{dz}(\hat{\bf
n}_2,z)dz}~.
\end{equation}

Given a map of CMB and QSO survey, the angular auto-correlation and
cross-correlation functions can be easily expressed in the harmonic
space:
\begin{eqnarray}
C^{\rm qT}(\theta)&\equiv&\left\langle\frac{\Delta{T}}{T}(\hat{\bf
n}_1)\delta_{\rm q}(\hat{\bf
n}_2)\right\rangle\nonumber\\
&=&\sum^{\infty}_{l=2}\frac{2l+1}{4\pi}C^{\rm qT}_lP_l[\cos(\theta)]~,\\
C^{\rm qq}(\theta)&\equiv&\left\langle\delta_{\rm q}(\hat{\bf
n}_1)\delta_{\rm q}(\hat{\bf
n}_2)\right\rangle\nonumber\\
&=&\sum^{\infty}_{l=2}\frac{2l+1}{4\pi}C^{\rm qq}_lP_l[\cos(\theta)]~,
\end{eqnarray}
where $\theta=|\hat{\bf n}_1-\hat{\bf n}_2|$ and the
auto-correlation and cross-correlation power spectra are given by:
\begin{eqnarray}
C^{\rm qT}_l&=&\frac{2}{\pi}\int{k^2dkP(k)I^{\rm ISW}_l(k)I^{\rm q}_l(k)}~,\\
C^{\rm qq}_l&=&\frac{2}{\pi}\int{k^2dkP(k)[I^{\rm q}_l(k)]^2}~,
\end{eqnarray}
where $P(k)$ is the matter power spectrum today and the functions
$I^{\rm ISW}_l(k)$ and $I^{\rm q}_l(k)$ are:
\begin{eqnarray}
I^{\rm ISW}_l(k)&=&-2\int{\frac{d\Phi(k)}{dz}j_l[k\chi(z)]dz}~,\\
I^{\rm q}_l(k)&=&\int{b_{\rm q}(z)\frac{dN}{dz}(z)\delta_{\rm
m}(k,z)j_l[k\chi(z)]dz}~, \label{jl}
\end{eqnarray}
where $j_l(x)$ are the spherical Bessel functions, and $\chi$ is the
comoving distance. In the following, we use the public package {\tt
CAMB${}_{-}$sources}\footnote{Available at
http://camb.info/sources/.} to calculate the theoretical angular
auto-correlation and cross-correlation functions.

\section{Quasar Catalog}
\label{qso}

We use the  SDSS DR6 quasar catalog released by
Ref.~\cite{richards09} (hereafter DR6-QSO).  This unique quasar
catalog contains about $N_{\rm q1}\approx 10^6$ objects with
photometric redshifts between $0.065$ and $6.075$, covering almost
all of the northern hemisphere of the galaxy plus three narrow
stripes in the southern, for a total area of $8417\,{\rm deg}^2$
($\sim20\%$ area of the whole sky). Photometrically selected QSOs
have become particularly important in the last few years due to the
higher selection efficiency reached, that has enabled their use for
meaningful statistical/cosmological analysis. The DR6-QSO data set
extends previous similar SDSS data sets with $\sim 95\%$ efficiency
\cite{richardsetal04,myersetal06}. The main differences are due to
the fact that DR6-QSO probes QSOs at higher redshift and also
contains putative QSOs flagged as to have ultra violet excess (UVX
objects). We refer the reader to Ref.~\cite{richards09} for a very
detailed description of the object selection with the non-parametric
Bayesian classification kernel density estimator (NBC-KDE)
algorithm.

\begin{figure}[t]
\begin{center}
\includegraphics[scale=0.45]{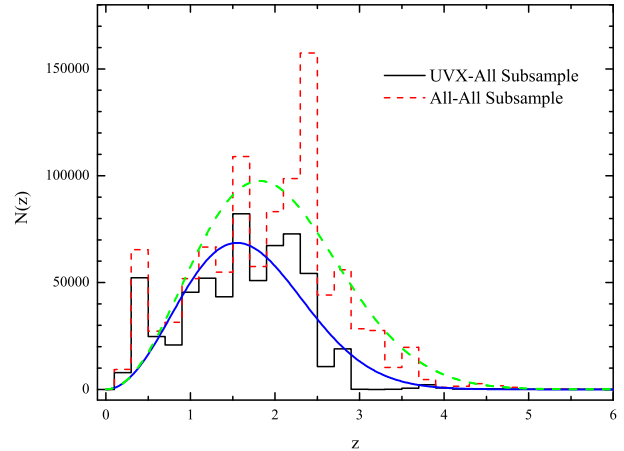}
\caption{Redshift distributions of the two quasar subsamples: black
solid line is for the ``UVX-All'' subsample while red dashed line is
for the ``All-All'' subsample. We also plot the theoretical redshift
distribution function $dN/dz$ of two subsamples: blue solid line and
green dashed line are for ``UVX-All'' and ``All-All'' subsample,
respectively.\label{fig1}}
\end{center}
\end{figure}

For our purposes we rely on the electronically published table that
contains only objects with the ``good'' flag with values within the
range $[0,6]$, and we label this as ``All-All'' subsample. The
higher the value, the more probable for the object to be a real QSO
(see Section 4.2 of Ref.~\cite{richards09} for details). As a
further more conservative criterion, we will present results for a
subset with ``uvxts=1'', i.e. objects clearly showing a UV excess
which should be a signature of a QSO spectrum. We are left with
$N_{\rm q2}\approx6\times10^5$ quasars and we refer to  this sample
as ``UVX-All'' subsample. This second choice is clearly more
conservative than the first one and thereby results should be
trusted more at a quantitative level.

For simplicity, we assume that the redshift distribution
$dN/dz$ of the DR6-QSO sample is approximated by the function:
\begin{equation}
\frac{dN}{dz}(z)=\frac{\beta}{\Gamma(\frac{m+1}{\beta})}\frac{z^m}{z^{m+1}_0}
\exp\left[-\left(\frac{z}{z_0}\right)^\beta\right]~,\label{reddis}
\end{equation}
where $m$, $\beta$ and $z_0$ are three free parameters, which are
dependent on the redshift distribution of quasar number density. For
the ``All-All'' subsample, we find that $m=2.00$, $\beta=2.20$, and
$z_0=1.90$. The mean redshift of this subsample is
$\bar{z}\sim1.80$. The ``UVX-All'' subsample has $m=2.00$,
$\beta=2.20$, $z_0=1.62$ and its mean redshift is $\bar{z}\sim1.49$.
The distributions are shown together with the redshift distribution
of the quasar number density in Fig.~\ref{fig1}. In this plot we do
not normalize the distributions to be unity.

Furthermore, we also take into account the magnification bias effect
which could be important for the SDSS QSOs. In the presence of
magnification bias, the relevant quantity entering Eq.~(\ref{jl}) is
the function $f(z)$  given by \cite{hoetal08}:
\begin{equation}
f(z)=b(z)\frac{dN}{dz}(z)+\int^{\infty}_zW(z,z')[\alpha(z')-1]\frac{dN}{dz'}dz'~,
\end{equation}
where $\alpha(z')$ is the slope of the number counts of the quasar
number density as a function of flux: $N(>F)\propto F^{-\alpha}$.
For simplicity, in our analysis we set $\alpha\equiv0.9$ in the
whole redshift region. Here in the flat universe the lensing window
function $W(z,z')$ is \cite{hoetal08}:
\begin{equation}
W(z,z')=\frac{3}{2}\Omega_{\rm m}H^2_0\frac{1+z}{cH(z)}\chi^2(z)
\left[\frac{1}{\chi(z)}-\frac{1}{\chi(z')}\right]~,
\end{equation}
where $\chi(z)=\int^z_0dz''/H(z'')$ is the radial comoving distance.

Besides the quasar sample in the whole redshift region, we are also
interested in the quasar subsamples in the low redshift $z<1.5$ and
the high redshift $z>1.5$. In the ``All-All'' subsample, there are
$N_{\rm lq1}\approx3.54\times10^5$ QSOs at low redshifts $z<1.5$
(``All-Low'') and $N_{\rm hq1}\approx6.57\times10^5$ QSOs at $z>1.5$
(``All-High''). The mean redshifts are $0.90$ and $2.28$,
respectively. For the ``UVX-All'' subsample, we are left with
$N_{\rm lq2}\approx2.87\times10^5$ (``UVX-Low'') and $N_{\rm
hq2}\approx3.21\times10^5$ (``UVX-High'') quasars. Their mean
redshifts are $0.90$ and $2.02$, respectively.

In these four cases, we also assume that the theoretical redshift
distributions are approximated by the function Eq.~(\ref{reddis}) in
the redshift region $[z_1,z_2]$. Furthermore, we widen these
redshift distributions by two one-tailed Gaussians when
$0\leq{z}<z_1$ and $z>z_2$ \cite{myersetal06}:
\begin{equation}
\frac{dN}{dz}(z)=[z_1,z_2]\exp\left[{-\frac{(z-[z_1,z_2])^2}{2\,\sigma^2}}\right]~,\label{widen}
\end{equation}
where $\sigma$ is the dispersion between photometric and
spectroscopic redshifts. This technique of widening by Gaussians is
consistent with determining $dN/dz$ from spectroscopic matches in
the photometric redshift bin \cite{myersetal06}. In
Ref.~\cite{myersetal06}, the authors measured this dispersion in the
different redshift bins and found $\sigma\sim0.2$. Therefore, in our
following calculations we simply set $\sigma=0.2$.

We have explicitly checked that these choices for the parameters
$\alpha$ and $\sigma$ have negligible impact on our final results.

\section{DR6-QSO Auto-Correlation Function}
\label{acf}

In this section we will use DR6-QSO sample to compute the QSO auto
correlation function (ACF).

\subsection{QSO Pixelation}
\label{pixel}

\begin{figure}[t]
\begin{center}
\includegraphics[scale=0.43]{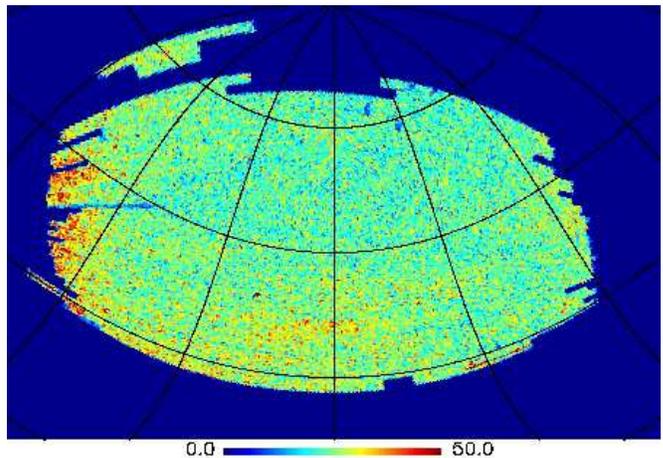}
\caption{Zoom on the SDSS DR6-QSO number density map with $N_{\rm
side}=128$ for the ``All-All'' subsample. This map is in celestial
coordinates. The grid spacing is $30^\circ$ and the center is at
${\rm RA}=180\,{\rm deg}$, ${\rm Dec}=37\,{\rm deg}$, corresponding
to $l=167.45$ and $b=75.3$ degrees in Galactic coordinates. RA is
increasing to the left.\label{fig2}}
\end{center}
\end{figure}

We pixelate the quasar maps using the HEALPix software
package\footnote{Available at http://healpix.jpl.nasa.gov/.}
\cite{healpix}. We use a relatively coarse resolution: $N_{\rm
side}=64$, corresponding to $N_{\rm pix}=49,152$ pixels with
dimensions $0.92^\circ\times0.92^\circ$. This resolution is
sufficient for the large scale correlations we are interested in. In
Fig.\ref{fig2}, for illustrative purposes, we show the quasar number
density map of the ``All-All'' subsample in celestial coordinates.
For the lowest resolution map, there are only $\sim20\%$ of these
pixels actually contain quasar sources.

In the low resolution map ($LN_{\rm side}=64$), there are many edge
pixels which are only partially filled by the QSOs. To account for
such an effect, following Ref.~\cite{gianna06}, we also use a higher
resolution pixelation map ($HN_{\rm side}= 512$) to estimate the
coverage fraction of each edge pixel.  In order to determine the
mask of the actual sky coverage of the DR6 survey, we generate a
random sample of galaxies using the DR6 database to ensure roughly
uniform sampling on the SDSS CasJobs website\footnote{Available at
http://casjobs.sdss.org/CasJobs/default.aspx/.}. By using a
sufficiently large number of random galaxies (between $13-20$
million) we can make sure to have a good sampling when these are
pixelated on the high resolution map.

Here, we choose  $LN_{\rm side}=64$ and $HN_{\rm side}=512$ for
the low and high resolution maps, respectively. We estimate the
coverage fraction of each low resolution pixel, $f_i$, as:
\begin{equation}
f_i=\frac{N^{\rm high}_{\rm mask}(i)}{64}~,
\end{equation}
where $N^{\rm high}_{\rm mask}(i)$ is the number of high resolution
pixels within the mask for each coarse pixel $i$, and for choice
made here there are $64=(HN_{\rm side}/LN_{\rm side})^2=(512/64)^2$
high resolution pixels in each coarse pixel. Using this method now
most of edge pixels are partially covered by the SDSS DR6 survey,
$0<f_i<1$.

We then correct the maps by dividing the observed number of quasars
in each coarse pixel by the fraction of the sky within the pixel
that was observed, yielding $n_i/f_i$. We use the higher resolution
to calculate the average number of quasars per coarse pixel,
$\bar{n}$:
\begin{equation}
\bar{n}=\frac{N_{\rm q}}{\sum{N^{\rm high}_{\rm
mask}(i)}}\times64=\frac{N_{\rm q}}{\sum{f_i}}~.
\end{equation}

We also use a higher low-resolution $LN_{\rm side}=128$ and a higher
high-resolution $HN_{\rm side}=1024$ to perform all the calculations
and find that our results are stable.

\subsection{ACF Estimator}
\label{acfest}

In order to measure the DR6-QSO ACF, we use the ACF estimator
$\hat{c}^{tt}(\theta)$, where the index $tt$ refers to the total
catalog (including possible stellar contaminations):
\begin{eqnarray}
\hat{c}^{tt}(\theta)&=&\frac{1}{N_{\theta}}\sum_{i,j}\frac{(n_i-f_i\bar{n})
(n_j-f_j\bar{n})}{\bar{n}^2}~,\label{acfestimator}\\
N_{\theta}&=&\sum_{i,j}f_if_j~,
\end{eqnarray}
where $f_i$ is the pixel coverage fraction, $n_i$ is the number of
quasar sources in each pixel, and $\bar{n}$ is the expectation value
for the number of objects in the pixel. The sum runs over all the
pixels with a given angular separation. For each angular bin
centered around $\theta$, $N_{\theta}$ is the number of pixels pairs
separated by an angle within the bin, weighted with the coverage
fractions.

This estimator is equivalent to the one used in Ref.~\cite{landy93}
in which the authors construct the ACF from the counts of data-data,
random-data and random-random number density pairs:
\begin{equation}
\hat{c}^{tt}(\theta)=\frac{QQ(\theta)+RR(\theta)-2QR(\theta)}{RR(\theta)}~,
\end{equation}
where $Q$ and $R$ denote the data point and random point,
respectively. In the following calculations, we also use this
estimator for cross-checking and find consistent results.

Because we pixelate the quasar map using a low resolution $N_{\rm
side}=64$, in which the pixel size is $55'$ \cite{healpix}, we use
$N_b=12$ angular bins in the range $1^\circ\leq\theta\leq12^\circ$
and a linear binning in our calculation. The choice of binning does
not affect the results significantly.

\subsection{Covariance Estimator}
\label{cov}

We estimate the covariance matrix of the data points using jackknife
resampling method \cite{scranton02}. This method is to divide the
data into $M$ patches, then create $M$ subsamples by neglecting each
patch in turn. These patches have roughly equal area. In practice,
we firstly list the whole set of pixels covered by the survey, and
then divide them into $M=30$ patches that do not have very similar
shape, but have roughly equal area (i.e. equal number of pixels).

The covariance estimator reads:
\begin{equation}
C_{ij}=\frac{M-1}{M}\sum^M_{k=1}\left[\hat{C}^{tt}_k(\theta_i)-\bar{C}^{tt}(\theta_i)\right]
\left[\hat{C}^{tt}_k(\theta_j)-\bar{C}^{tt}(\theta_j)\right],\label{covest}
\end{equation}
where $\hat{C}^{tt}_k(\theta_i)$ are the observed ACF of the $M$
subsamples in the $i$-th angular bin and $\bar{C}^{tt}(\theta_i)$
are the mean ACF over $M$ realizations. The diagonal part of these
matrices gives the variance of the ACF in each bin
$C^k_{ii}=\sigma^2_i$, while the off-diagonal part represents the
covariance between the angular bins. We also change the number of
patches $M$ and verify that the covariance matrix is stable.  We
refer the readers to Ref.~\cite{gianna08} for a more extensive
comparison between several covariance estimators.

\subsection{Stellar Contamination}
\label{star}

Although the SDSS DR6-QSO catalog we use has very high efficiency in
the selection algorithm, stars are point-like sources that
inevitably contaminate the catalog. An estimate of the level of
stellar contamination can be computed using the fact that the
correlations properties of stars are very different from those of
QSOs. In particular, a nearly flat contribution, up to large angular
scales (that correspond to small physical Galactic distances) is
expected from stars of our own galaxy.

If we substitute $n_i\rightarrow an^q_i+(1-a)n^s_i$, where $a$ is
the efficiency of the quasar catalog, the ACF estimator of
Eq.~(\ref{acfestimator}) becomes:
\begin{equation}
\hat{c}^{tt}(\theta)=
a^2\hat{c}^{qq}(\theta)+(1-a)^2\hat{c}^{ss}(\theta) +
\epsilon(\theta)~,\label{crossterm}
\end{equation}
where $\hat{c}^{qq}(\theta)$ and $\hat{c}^{ss}(\theta)$ are the
intrinsic ACF of QSOs and stars, respectively, and
$\epsilon(\theta)$ is a tiny offset arising from cross-terms
\cite{myersetal06}:
\begin{eqnarray}
\epsilon(\theta)&=&2(a-a^2)\left(\frac{QS(\theta)+RR(\theta)-QR(\theta)-SR(\theta)}{RR(\theta)}\right),\nonumber\\
&=&\frac{2(a-a^2)}{N_\theta}\sum_{i,j}\frac{(n^q_i-f_i\bar{n})
(n^s_j-f_j\bar{n})}{\bar{n}^2}~.
\end{eqnarray}
If the efficiency of the catalog is high enough, i.e. $a>90\%$,
this cross-term $\epsilon(\theta)$ should be close to zero and can
be neglected safely in the analysis. Otherwise, $\epsilon(\theta)$
will become large and is comparable to the intrinsic ACF of stars
$\hat{c}^{ss}(\theta)$. Thus, in order to correctly determine the
efficiency of the catalog with a large stellar contamination, such
as the ``All-All'' subsample, we have to take this term into
account. We will show this effect in the following sections.

\begin{figure}[t]
\begin{center}
\includegraphics[scale=0.47]{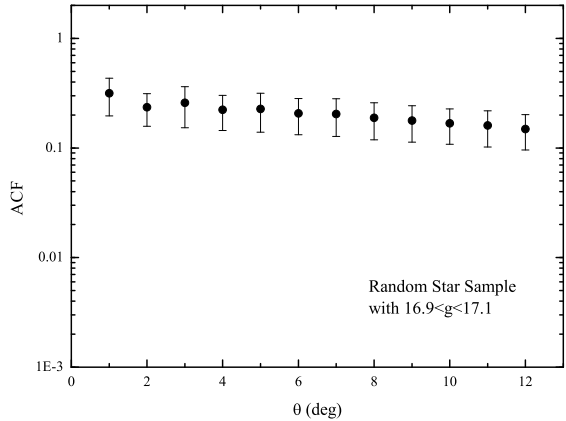}
\caption{The auto-correlation function of star catalog
$\hat{c}^{ss}(\theta)$. And the errors are also derived from the
jackknife method.\label{fig3}}
\end{center}
\end{figure}

To compute the stellar contamination, we extract a large number
($\sim 8\times10^4$) of stars in SDSS DR6 survey in the magnitude
range $16.9<g<17.1$ from the CasJobs website and compute the ACF of
stars $\hat{c}^{\rm ss}(\theta)$ using Eq.~(\ref{acfestimator}). In
Fig.~\ref{fig3} we show the stellar ACF and its errors which are
also measured with the jackknife method. We can find that the
stellar contamination will dominate the ACF of the total catalog at
the largest scales where the ACF of QSOs $\hat{c}^{\rm
qq}(\theta)\rightarrow0$.  The star contribution is indeed not
perfectly flat and retains a small slope even at large angles,
dropping from $\hat{c}^{\rm ss}(1^\circ)\sim0.3$ to $\hat{c}^{\rm
ss}(12^\circ)\sim0.15$, which is consistent with the result of
Ref.~\cite{myersetal06}, showing that probably our star catalog is
also contaminated at some level by QSOs. However, the contribution
at large angular scales could be robustly estimated and removed.

\subsection{Systematic Errors}
\label{sys}

Several systematics effects, including the galactic extinction by
dust, sky brightness, number of point sources and poor seeing, could
potentially affect both the observed ACF and CCF. These systematics
are investigated in detail in
Refs.~\cite{gianna06,hoetal08,myersetal06}. We also checked for
their contribution in our calculation and in particular we consider
extinction and point sources contamination, that are believed to
affect most the measured correlations.

\begin{figure}[t]
\begin{center}
\includegraphics[scale=0.47]{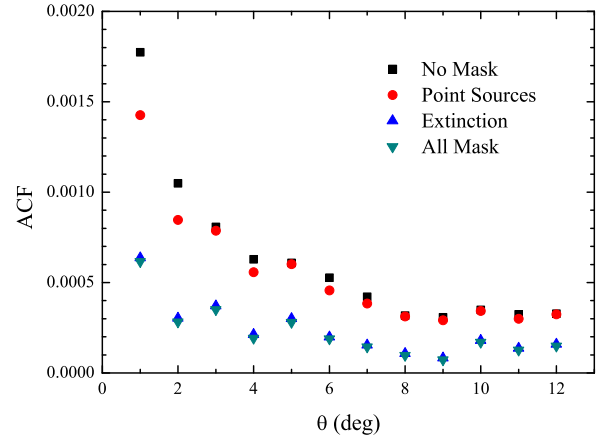}
\caption{The auto-correlation function of quasar ``UVX-All''
subsample measured for all the sample, for a single foreground mask
and for all masks joint. No error bars are reported.\label{fig4}}
\end{center}
\end{figure}

\begin{figure}[t]
\begin{center}
\includegraphics[scale=0.47]{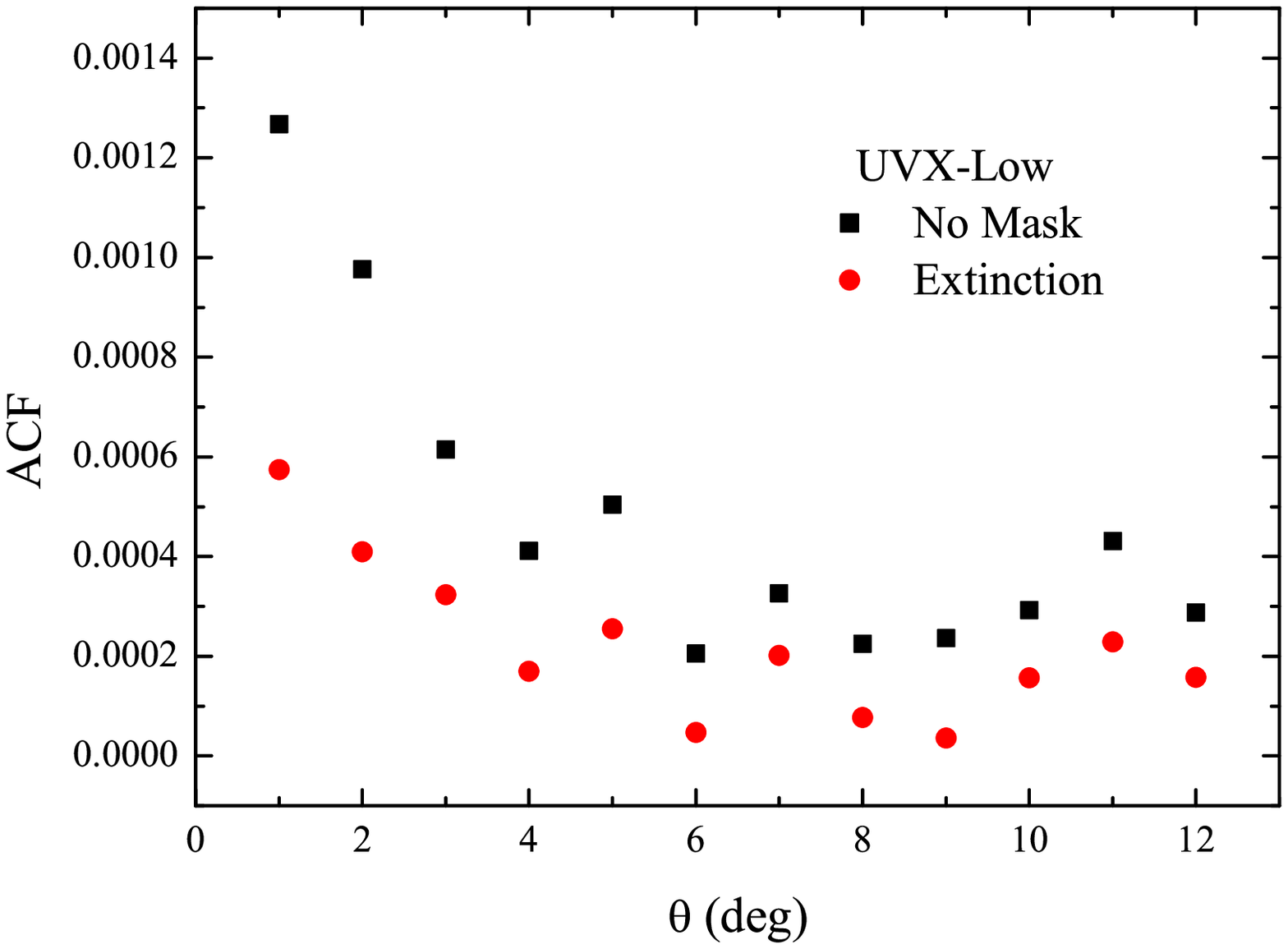}
\includegraphics[scale=0.47]{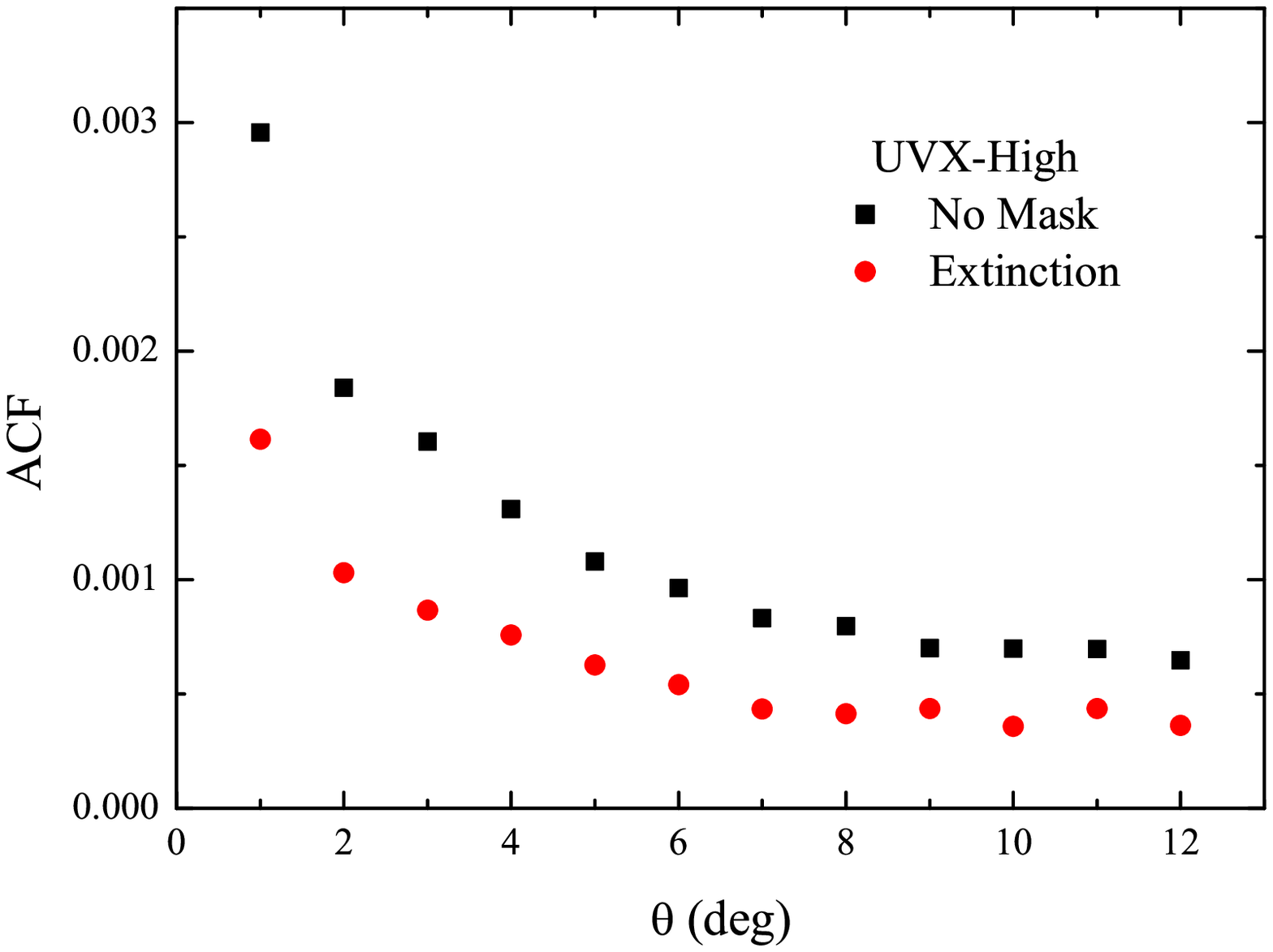}
\caption{The auto-correlation function of quasar ``UVX-Low'' and
``UVX-High'' subsamples measured for the extinction mask. No error
bars are reported.\label{figext1}}
\end{center}
\end{figure}

Basically we recompute the ACF by considering only objects within
the pixels with $A_{\rm g}<0.18$ (the g-band Galactic extinction)
for the reddening, while for the point sources we remove the pixels
with more than twice average number density of sources,
$n_i<2\,\bar{n}$. These masks will remove about $\sim20\%$ of the
considered area. Among these two systematics we find that the
extinction have the major effect on the derived ACF and we plot the
observed ACF with and without these foreground masks for the
``UVX-All'' subsample in Fig.\ref{fig4}. We also plot the effect of
extinction on the ACF for the ``UVX-Low'' and ``UVX-High''
subsamples in Fig.\ref{figext1}. Therefore, we will use the
extinction mask $A_{\rm g}<0.18$ only in our following analysis. Our
results are in quantitative agreement with those of
Ref.~\cite{gianna06} based on SDSS DR4 quasar catalog.

\begin{figure*}[t]
\begin{center}
\includegraphics[scale=0.29]{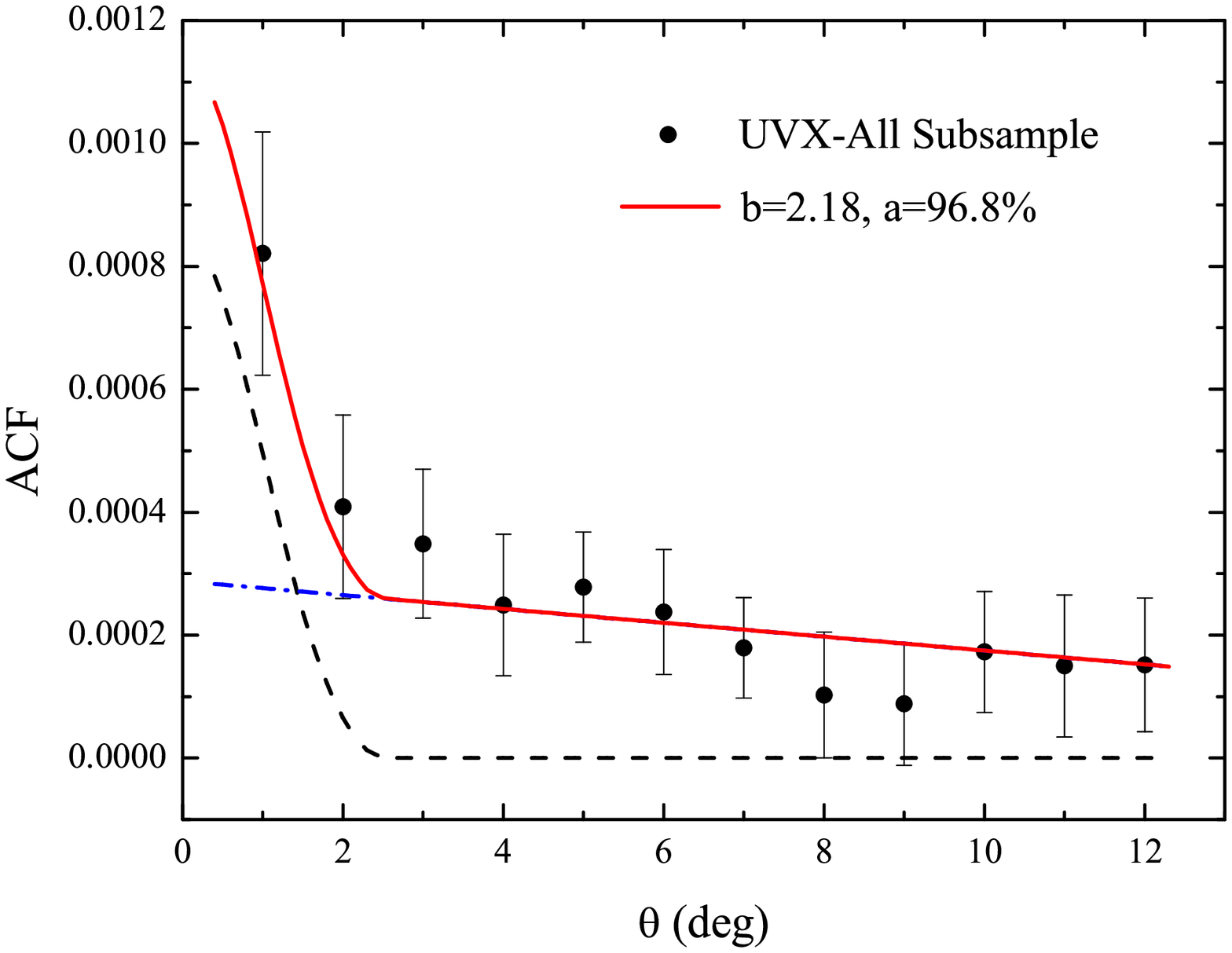}
\includegraphics[scale=0.29]{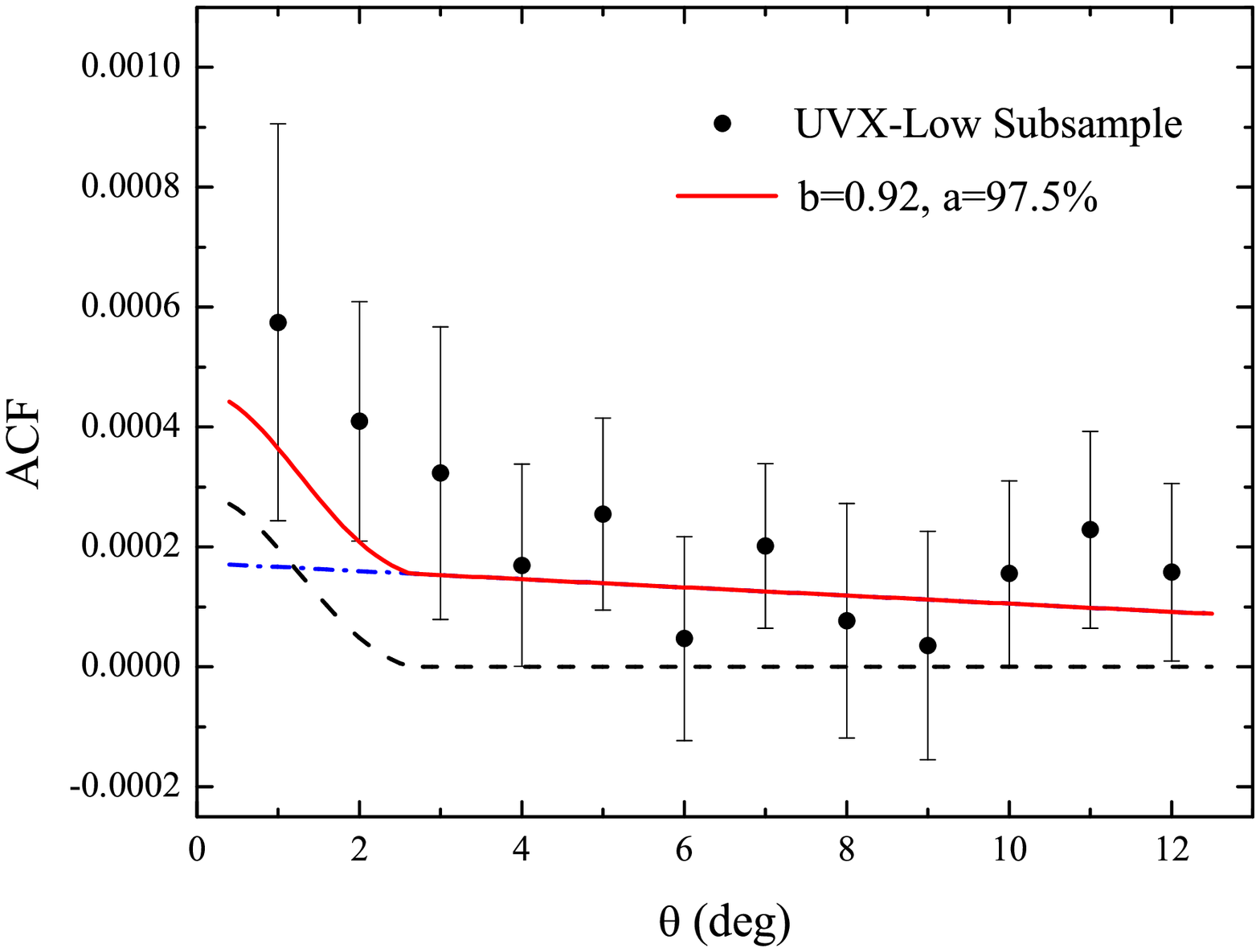}
\includegraphics[scale=0.29]{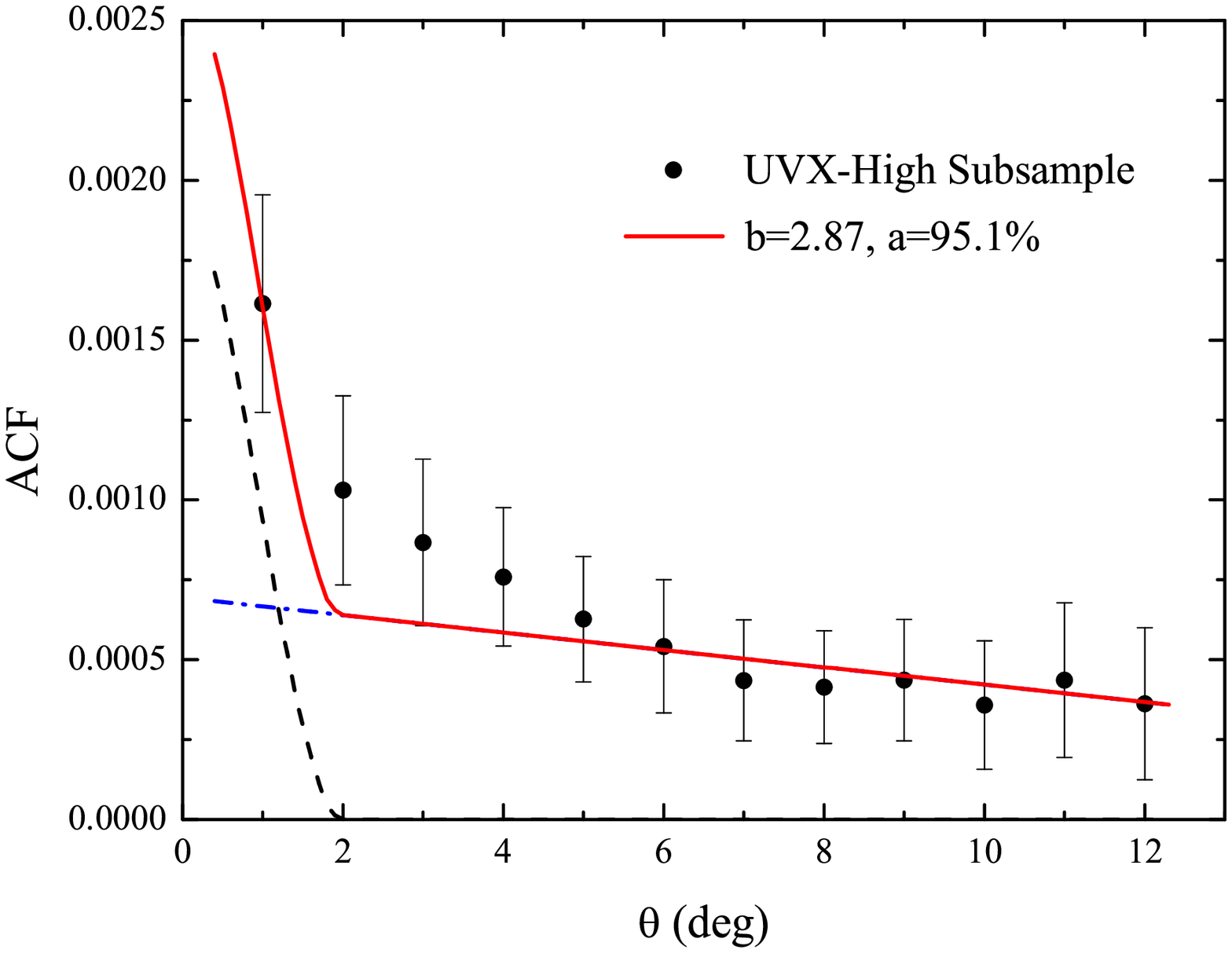}
\includegraphics[scale=0.29]{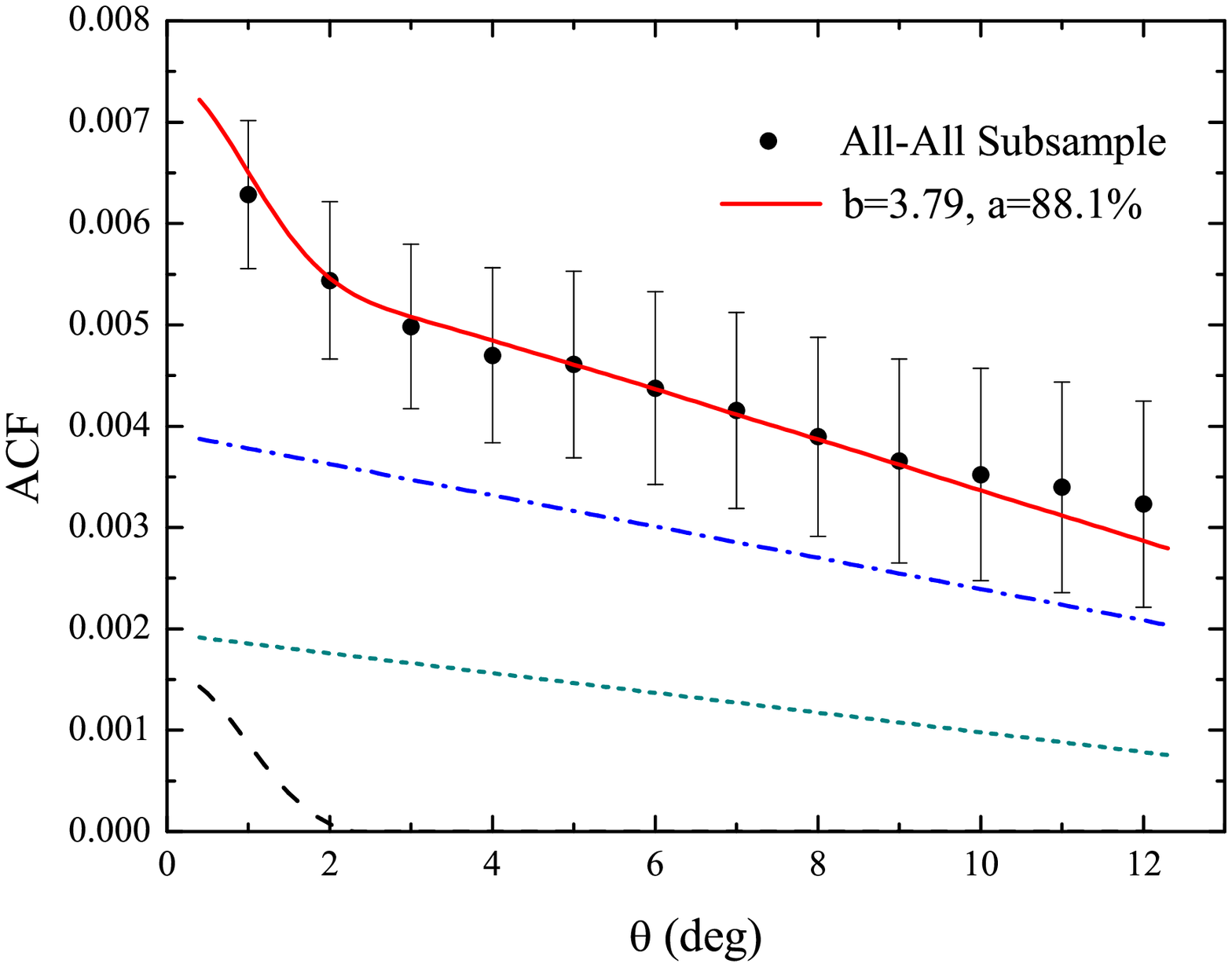}
\includegraphics[scale=0.29]{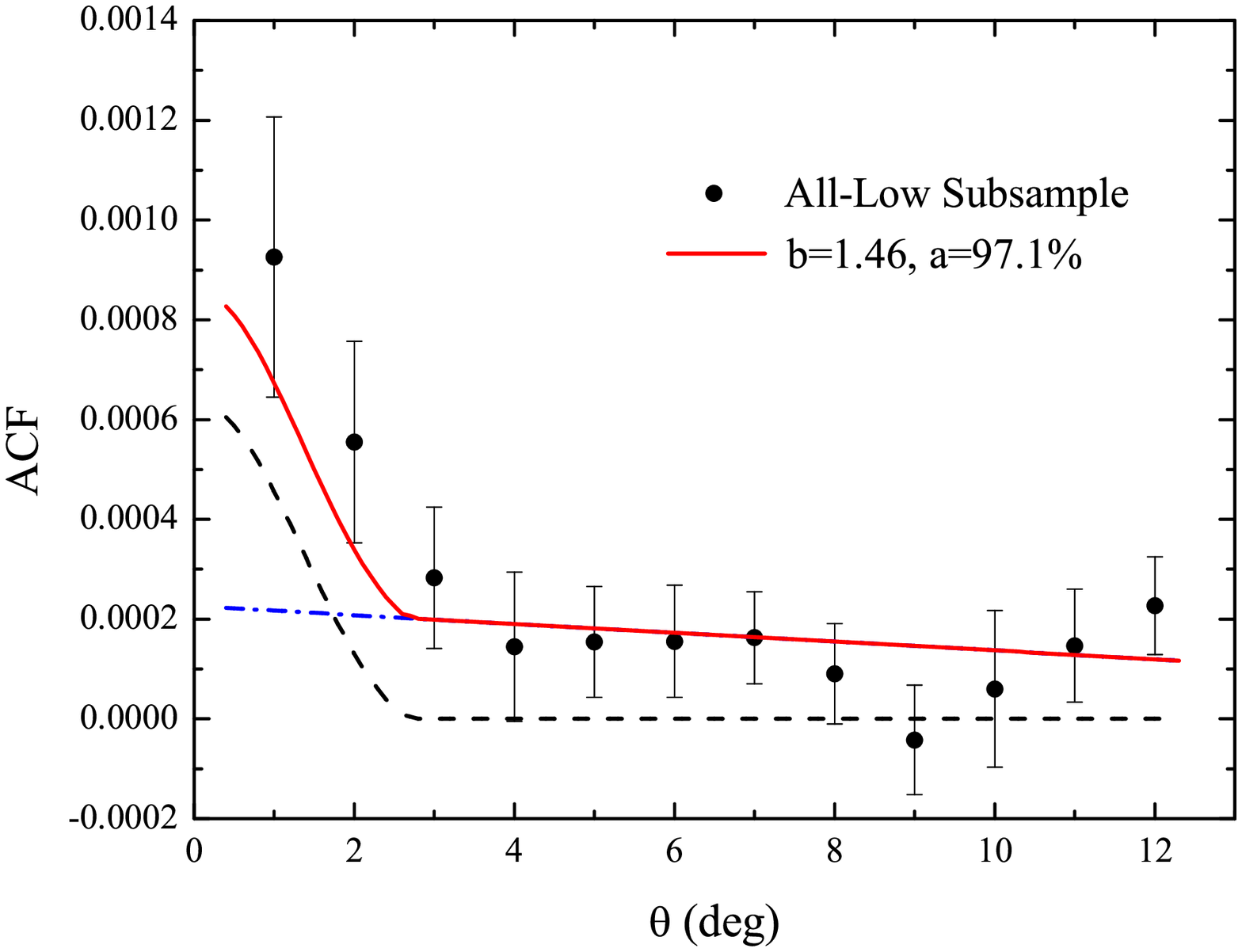}
\includegraphics[scale=0.29]{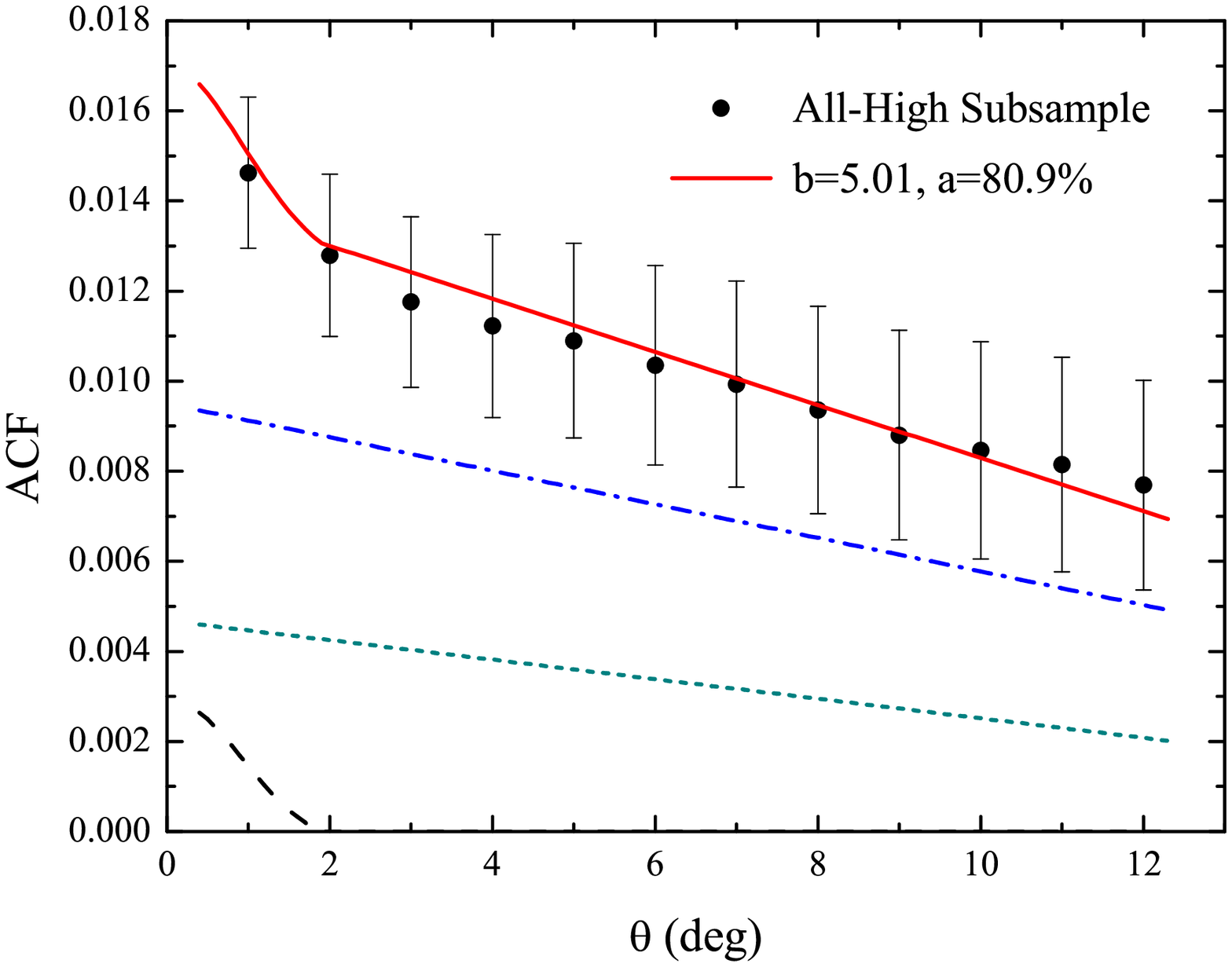}
\caption{The observed auto-correlation function (ACF) of different
quasar subsamples. Top three panels are for ``UVX-All'', ``UVX-Low''
and ``UVX-High'' subsamples. Bottom three panels are for
``All-All'', ``All-Low'' and ``All-High'' subsamples. All errors are
estimated with jackknife method. The red solid lines are the
theoretical predictions from the best fit model of WMAP five-year
data, while the black dashed lines and the blue dash-dot lines are
its component (theoretical quasars ACF and stellar contamination).
We also plot the contributions from the cross term
$\epsilon(\theta)$ (cyan short dashed lines) in the ``All-All'' and
``All-High'' subsamples in which the efficiencies become small.
\label{fig5}}
\end{center}
\end{figure*}

\subsection{Auto-Correlation Functions}
\label{calacf}

In this subsection, we summarize the whole calculations of DR6-QSO
ACF. For our purpose we construct six different quasar subsamples
from the SDSS DR6-QSO catalog. Firstly we pixelate the quasar maps
using HEALPix software and remove the $\sim20\%$ highest
contaminated pixels with $A_{\rm g}>0.18$. Next we use the ACF
estimator Eq.~(\ref{acfestimator}) and the covariance estimator
Eq.~(\ref{covest}) to calculate the ACF and its covariance matrix
for each quasar subsample. Finally we consider the contribution of
stellar contamination and also the cross terms when the efficiency
is not high enough. Note that in the following analysis we  take in
account the window function $w(\theta)$ associated with our
pixelation.

In Fig.~\ref{fig5} we plot the observed ACF for these six different
quasar subsamples. We find that for the ``UVX-All'' subsample the
values of ACF are consistent with previous works
\cite{gianna06,gianna08}. However, if we consider the high redshift
subsample, the observed ACF becomes large. The reason is that in
this case the subsample has a higher mean redshift. At this high
mean redshift, the bias and stellar contamination will become larger
and produce a larger ACF (the efficiency is smaller). Moreover, due
to their higher mean redshift, larger bias and smaller efficiency,
three ``All-$\times\times\times$'' subsamples also have larger
observed ACFs than the corresponding three
``UVX-$\times\times\times$'' subsamples. In Fig.\ref{fig5} we show
the theoretical ACFs for different cases. Obviously all the plots
depend on the chosen cosmology. Here we use the best fit model of
WMAP five-year data \cite{Komatsu:2008hk}: $\Omega_{\rm
b}h^2=0.02267$, $\Omega_{\rm c}h^2=0.1131$, $\tau=0.084$, $h=0.705$,
$n_{\rm s}=0.96$ and $A_{\rm s}=2.15\times10^{-9}$ at $k=0.05\,{\rm
Mpc}^{-1}$. For the other two free parameters, bias and efficiency,
we choose the best fit values of Table II, obtained from our
calculations in the $\Lambda$CDM framework that will be shown in the
following sections. We find that there is good agreement between the
theory from the WMAP5 best fit model and the observed ACF in each
quasar subsample.

\section{QSO-CMB Cross-Correlation Function}
\label{ccf}

\begin{figure}[t]
\begin{center}
\includegraphics[scale=0.43]{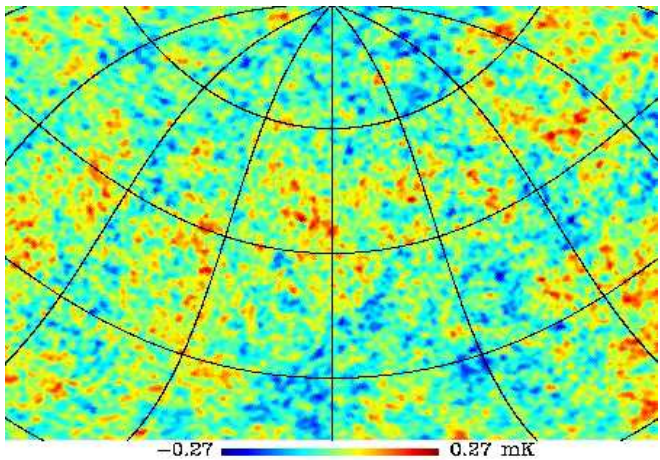}
\caption{WMAP Internal Linear Combination map with $N_{\rm
side}=128$ in celestial coordinates zoomed on the same region probed
by SDSS DR6-QSOs. The grid spacing and center are the same as in
Fig.\ref{fig2}. \label{fig6}}
\end{center}
\end{figure}

\begin{figure*}[t]
\begin{center}
\includegraphics[scale=0.29]{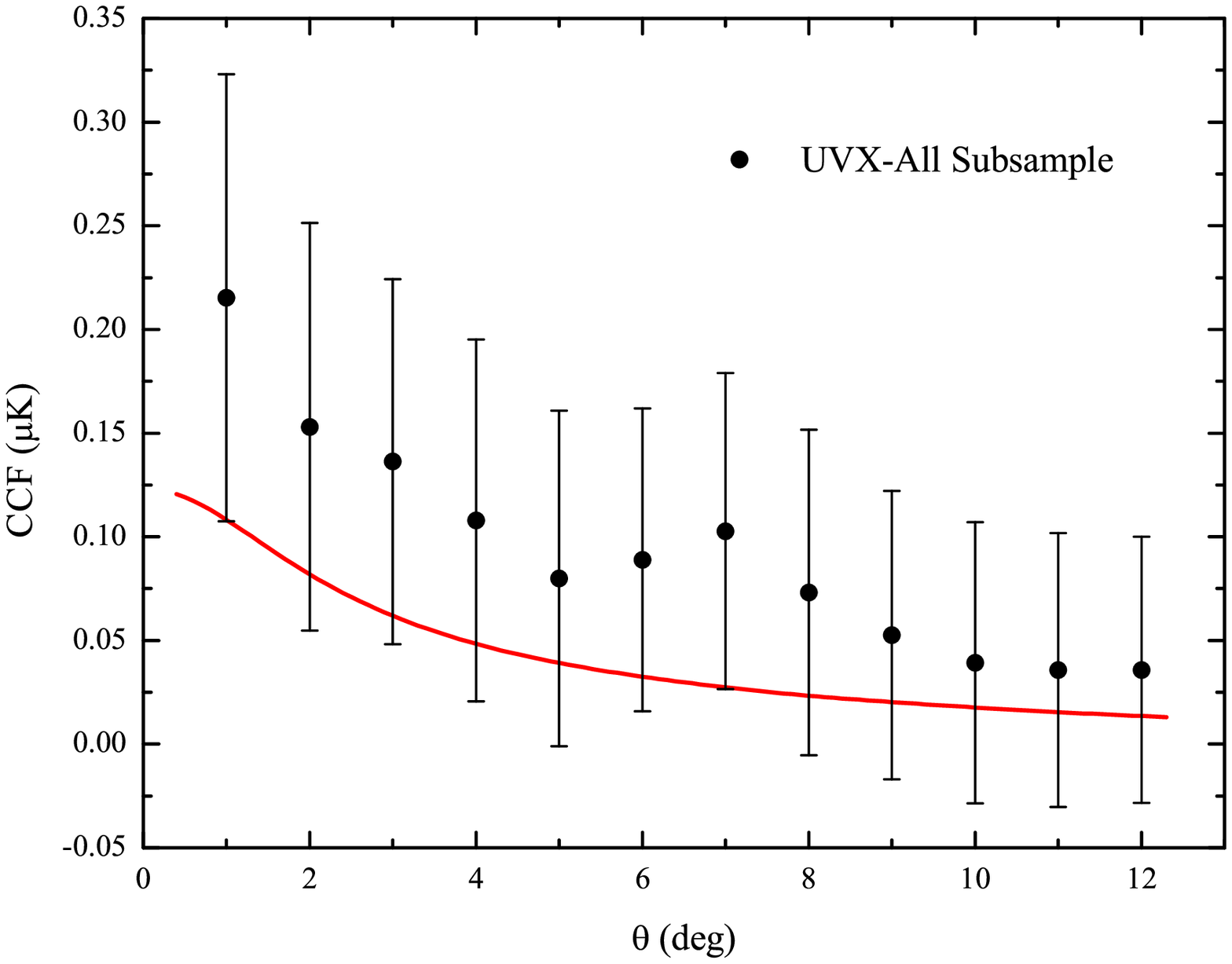}
\includegraphics[scale=0.29]{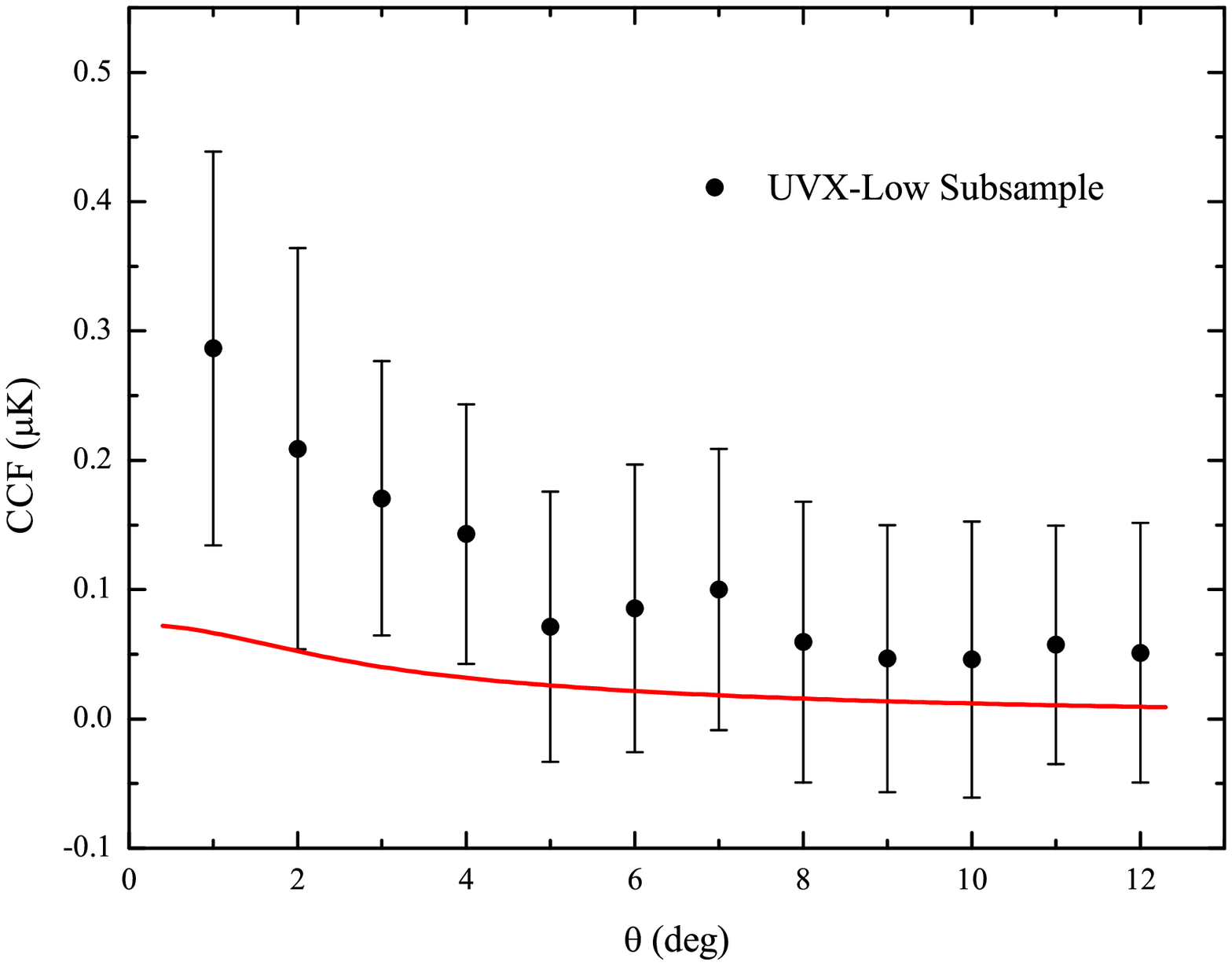}
\includegraphics[scale=0.29]{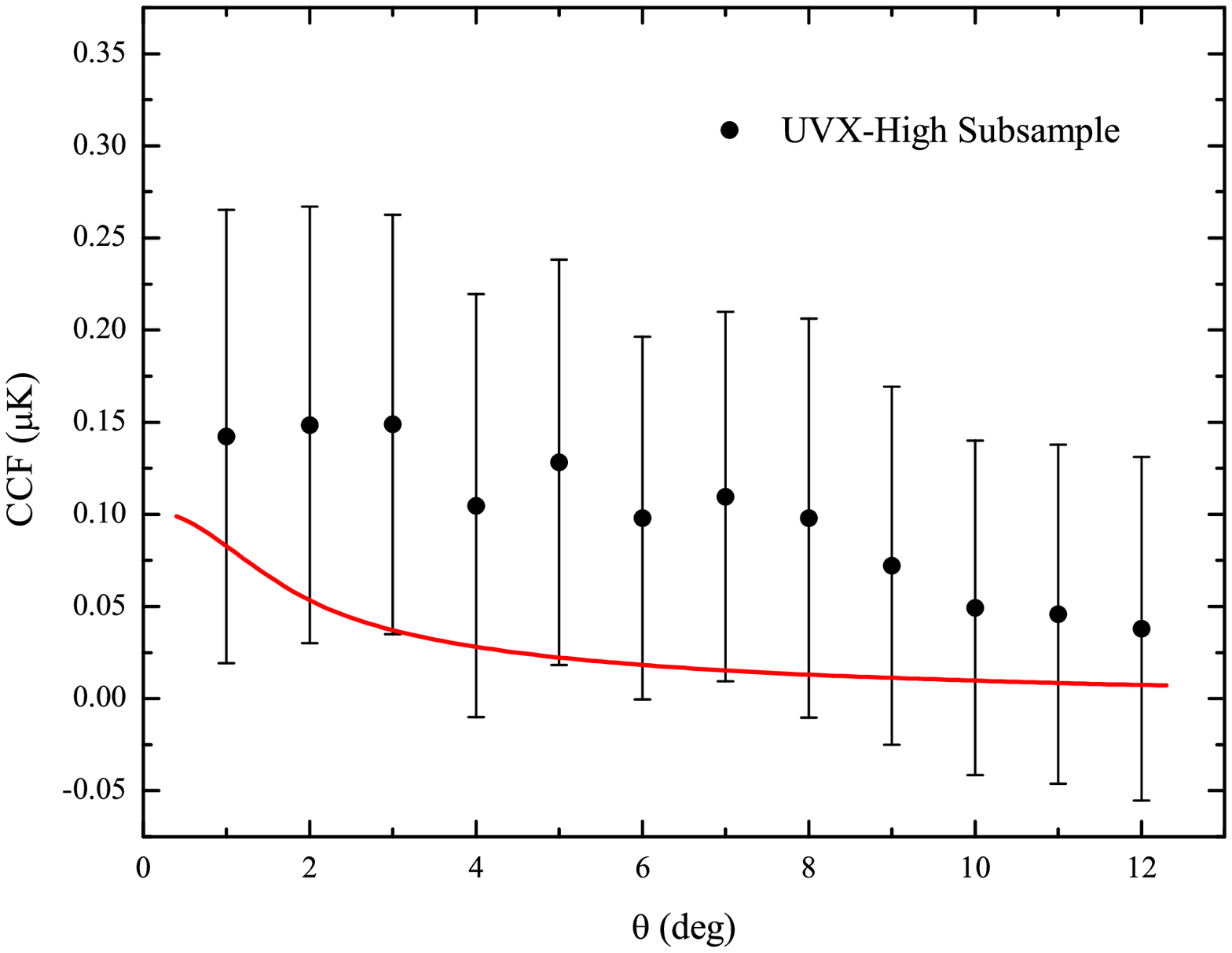}
\includegraphics[scale=0.29]{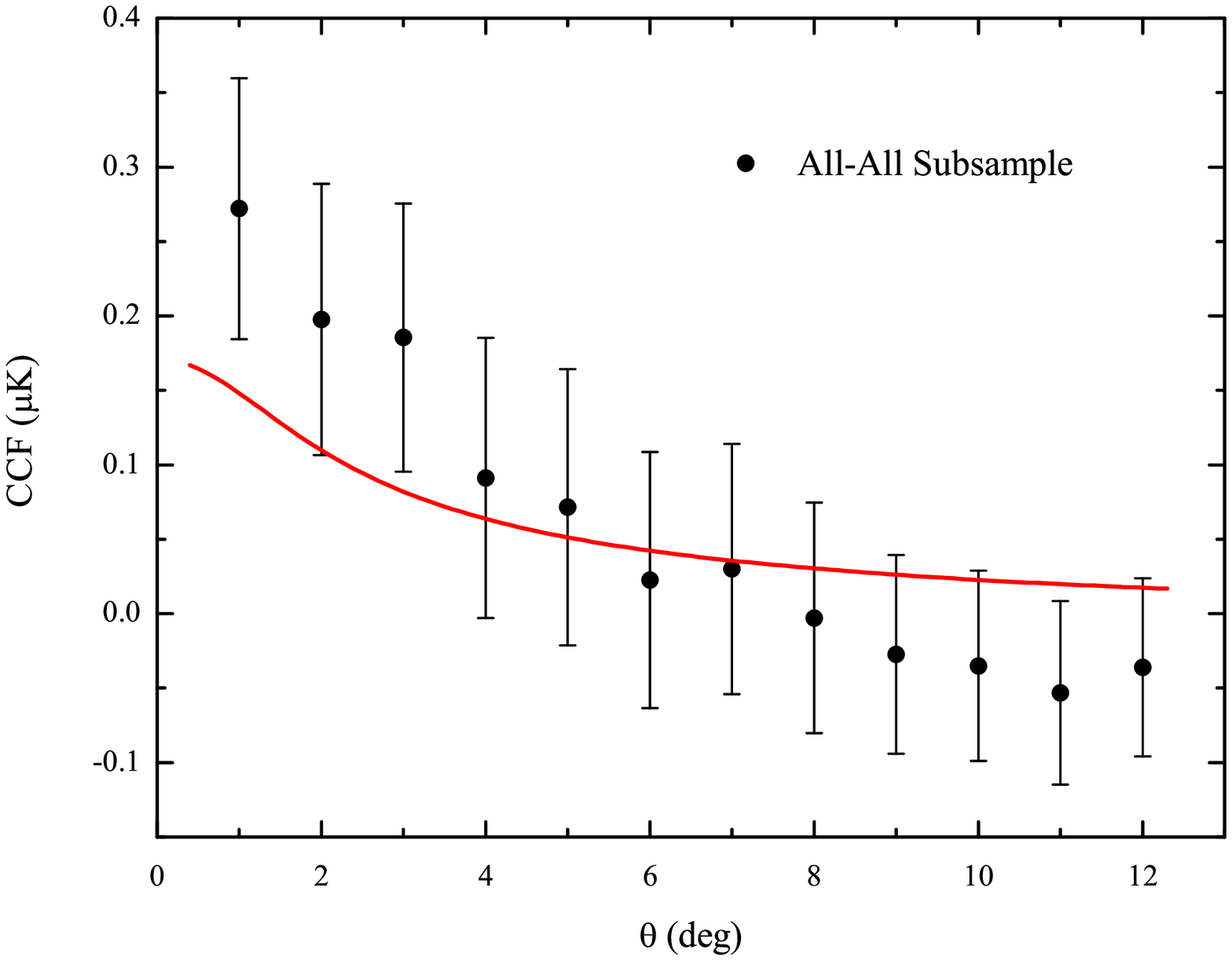}
\includegraphics[scale=0.29]{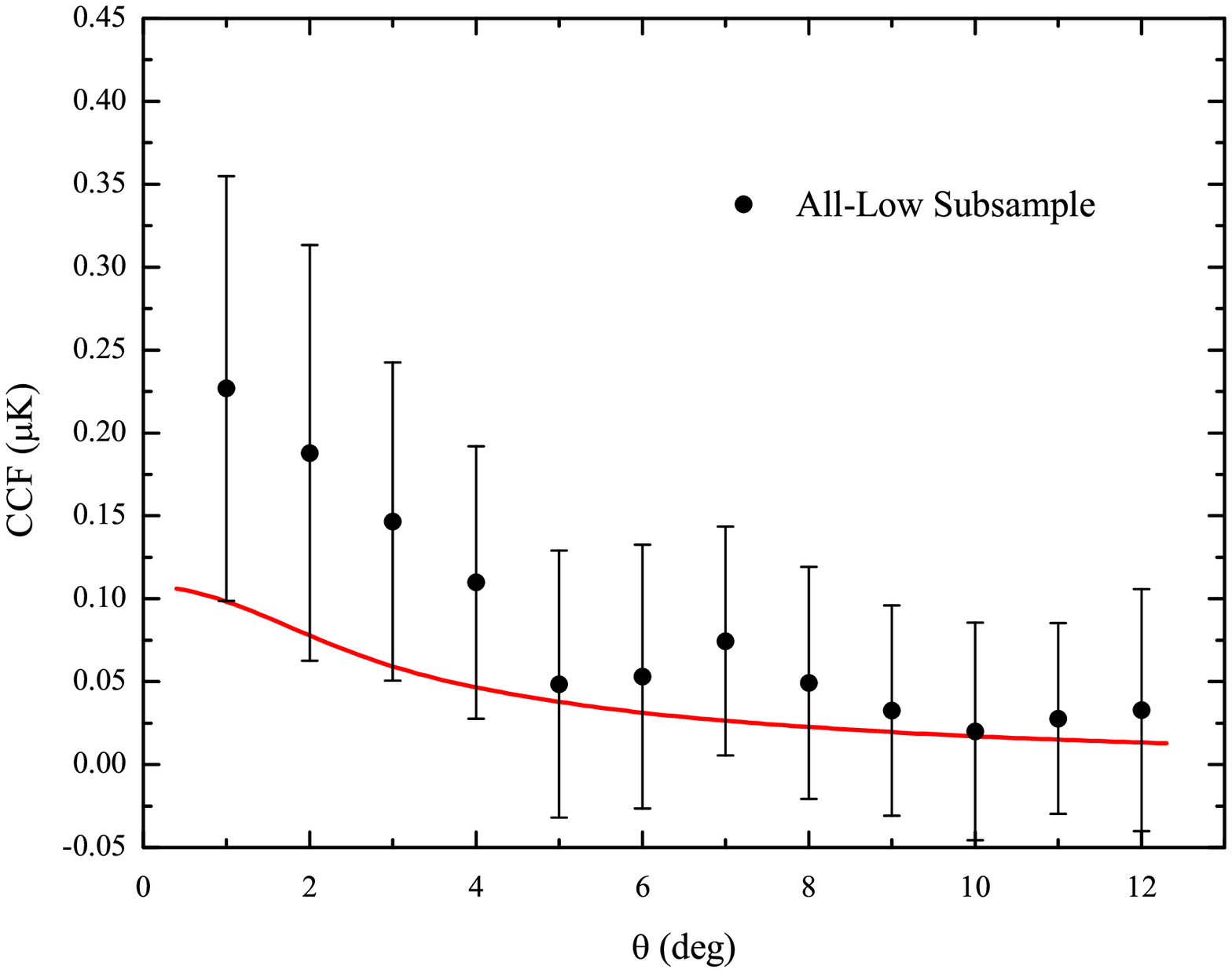}
\includegraphics[scale=0.29]{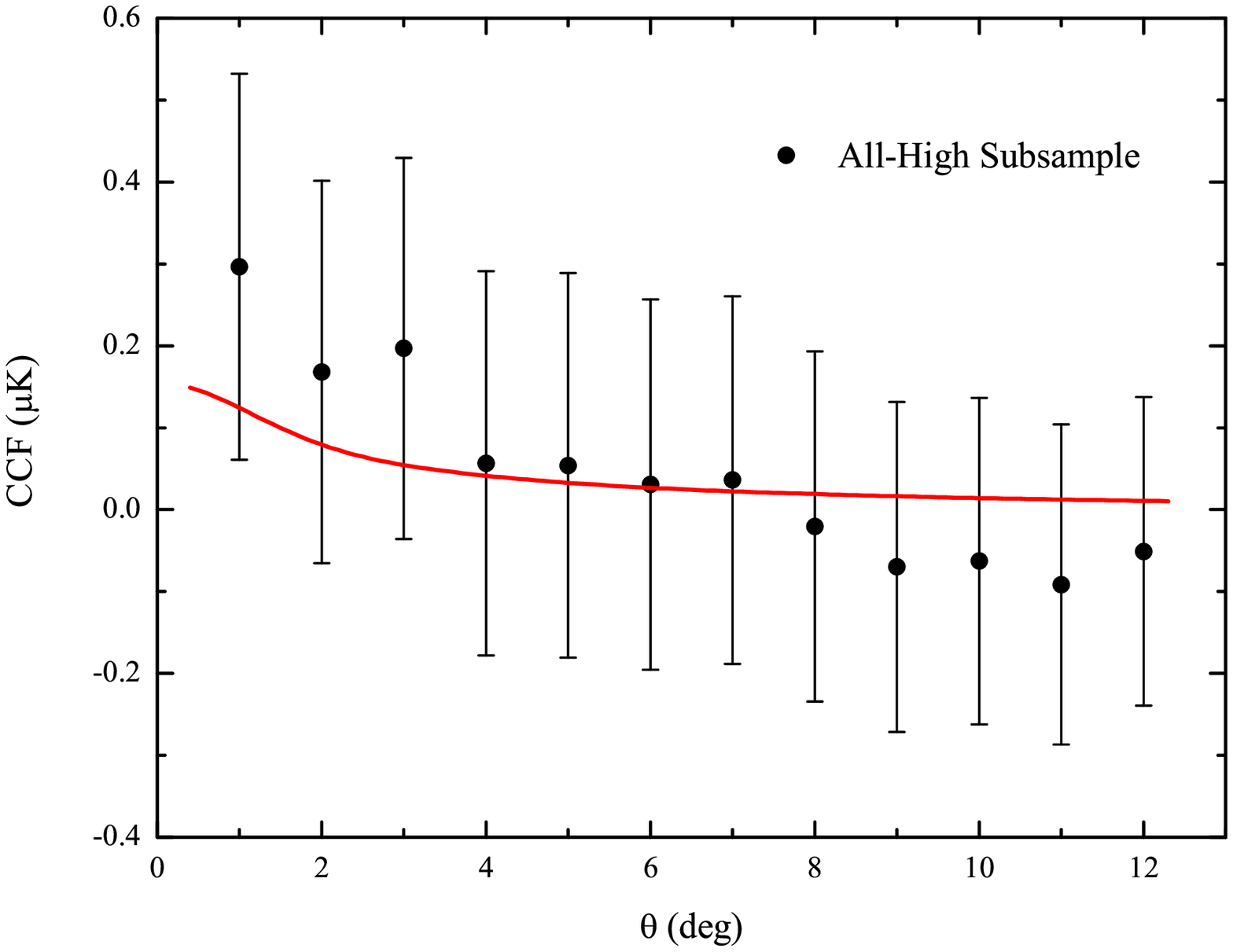}
\caption{The observed cross-correlation function (CCF) between
different quasar subsamples and WMAP ILC map. Top three panels are
for ``UVX-All'', ``UVX-Low'' and ``UVX-High'' quasar subsamples.
Bottom three panels are for ``All-All'', ``All-Low'' and
``All-High'' quasar subsamples. All errors are jackknife estimated.
The red solid lines are the theoretical predictions from the best
fit model of WMAP five-year data.\label{fig7}}
\end{center}
\end{figure*}

\begin{figure}[t]
\begin{center}
\includegraphics[scale=0.47]{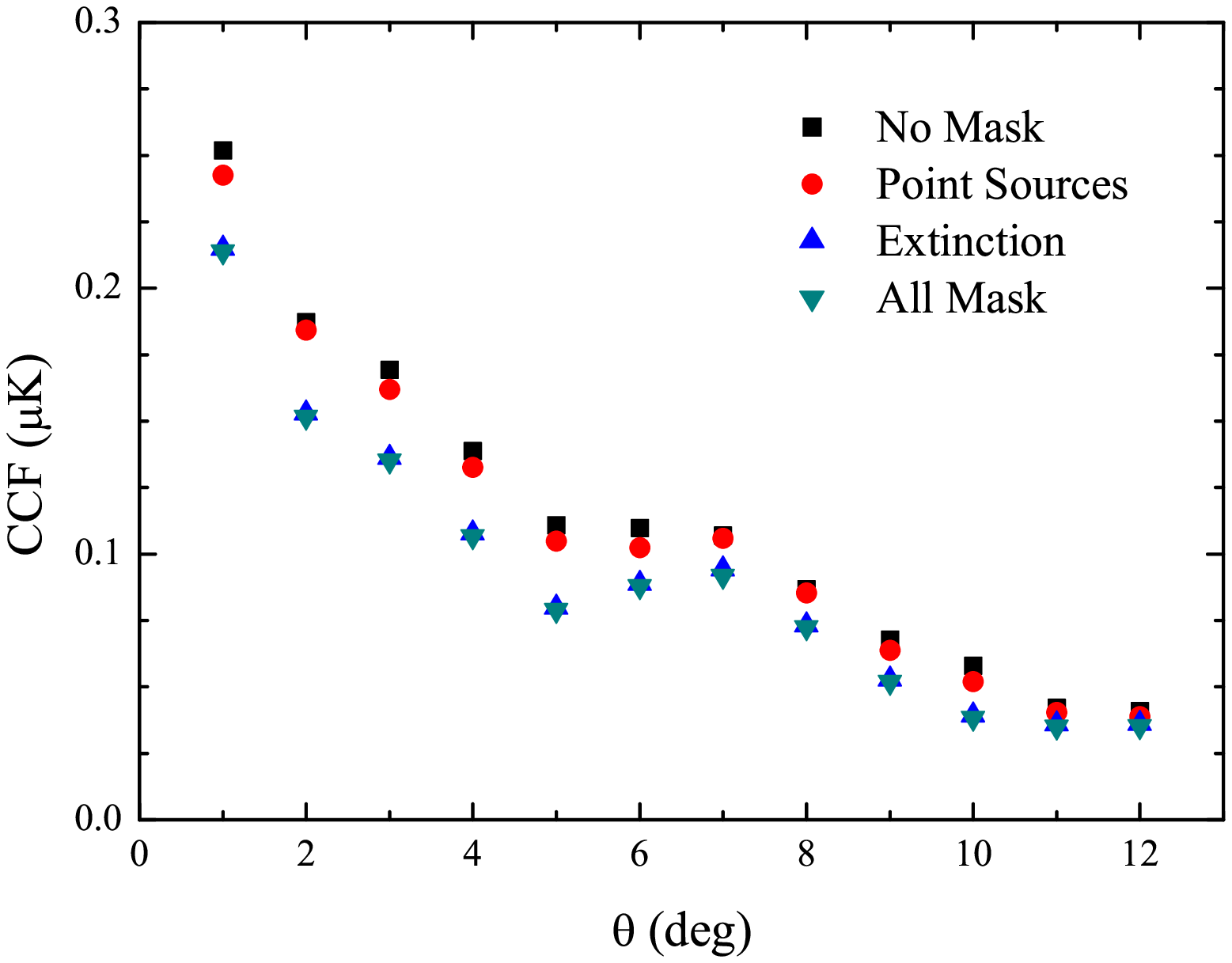}
\caption{The cross-correlation function of quasar ``UVX-All''
subsample measured for all the sample, for a single foreground mask
and for all masks joint. No error bars are reported.\label{figccf}}
\end{center}
\end{figure}

\begin{figure}[t]
\begin{center}
\includegraphics[scale=0.47]{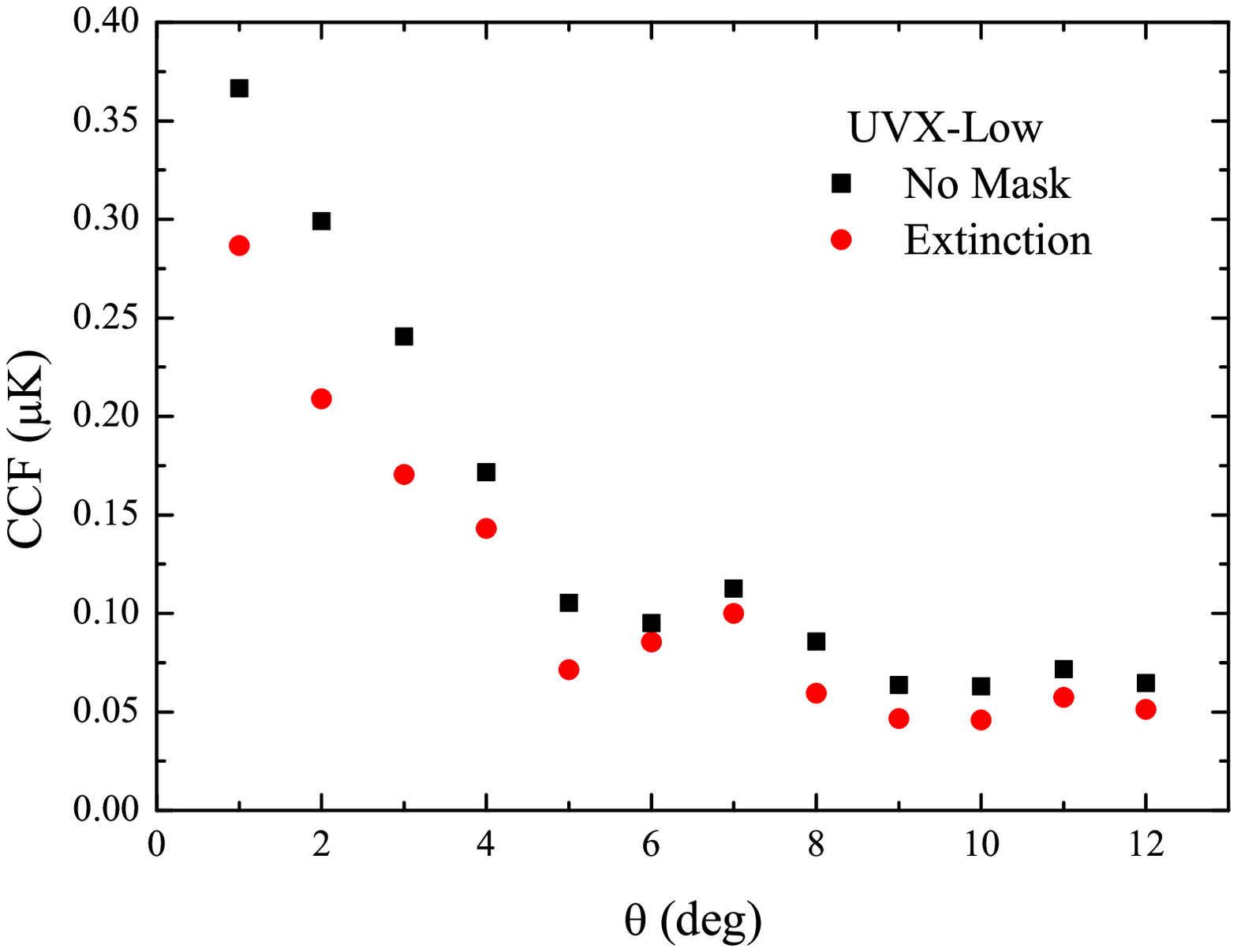}
\includegraphics[scale=0.47]{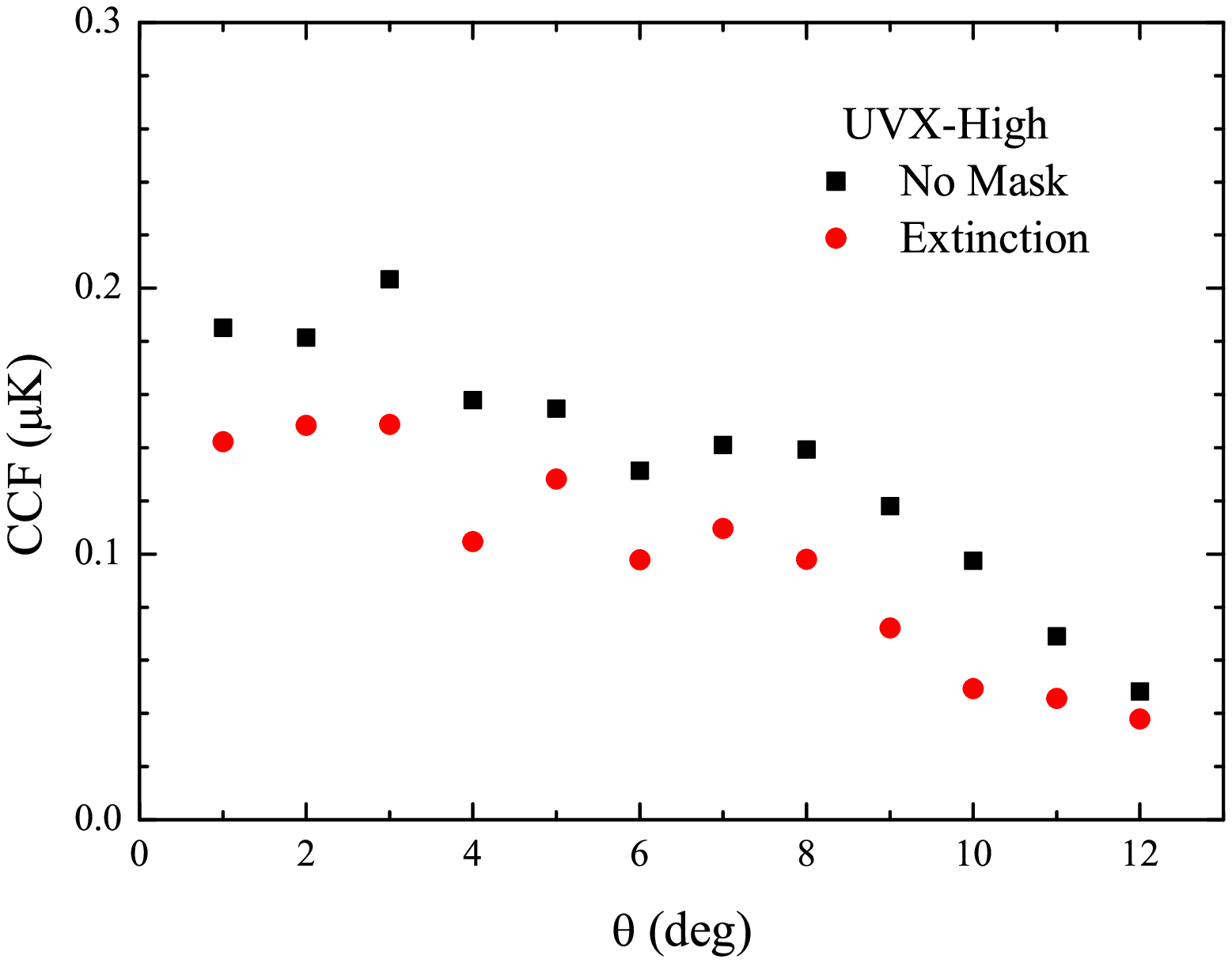}
\caption{The cross-correlation function of quasar ``UVX-Low'' and
``UVX-High'' subsamples measured for the extinction mask. No error
bars are reported.\label{figext2}}
\end{center}
\end{figure}

For the cross-correlation analysis, we use the WMAP Internal Linear
Combination (ILC) map derived from the five-year WMAP data, with
$HN_{\rm side}=512$ provided by the WMAP team \cite{hinshaw09},
shown in Fig.\ref{fig6}, in the same region as the SDSS DR6-QSO
data. This ILC map was already built to minimize the Galactic and
other foreground contaminations. In our calculations, we down-grade
the ILC map to the low resolution $LN_{\rm side}=64$. For the WMAP
mask, we use the ``KQ75'' mask \cite{gold09} corresponding roughly
to the ``Kp0" cut in the three-year data release. This mask map has
also $N_{\rm side}=512$.  We down-grade this map to the low
resolution and set the weight $w^{T}=0$ for all pixels including at
least one masked high resolution pixel \cite{raccanellietal08}.

To measure the CCF between the SDSS DR6-QSO number density map
and the WMAP ILC map, we rely on the following estimator:
\begin{equation}
\hat{c}^{\rm tT}(\theta)=\frac{1}{N_{\theta}}\sum_{i,j}(T_i-\bar{T})\frac{n_j-f_j\bar{n}}{\bar{n}}~,
\end{equation}
where $T_i$ is the CMB temperature in the $i$-th pixel and $\bar{T}$
is the mean (monopole) value for the CMB temperature in the unmasked
area. $N_{\theta}$ is the number of pixels pairs, which has been
defined in the previous Section. We also use $N_b=12$ angular bins
in the range $1^\circ\leq\theta\leq12^\circ$ and a linear binning in
our calculation. We also check the contributions from the extinction
and point sources contamination in Fig.\ref{figccf}, and find that
the extinction have the major effect on the CCF. We also plot the
effect of extinction on the CCF for the ``UVX-Low'' and ``UVX-High''
subsamples in Fig.\ref{figext2}. Therefore, we apply the reddening
mask with $A_g<0.18$ to be consistent with the ACF measurements.

We also consider the possible contributions from stars in the CCF
calculations. The stellar contamination has to be subtracted from
the total CCF and the QSO-temperature correlation that will be
compared with the theoretical models becomes:
\begin{equation}
\hat{c}^{\rm qT}(\theta)=\frac{\hat{c}^{\rm tT}(\theta)-(1-a)\hat{c}^{\rm sT}(\theta)}{a}~,
\end{equation}
where the star-temperature cross-correlation has to be also
estimated. 

The covariance matrices are also calculated by using jackknife
resampling method, Eq.~(\ref{covest}). We firstly list all the
pixels covered by the survey in the quasar map, and divide them into
$M=30$ patches. Then we create $M$ subsamples by neglecting each
patch in turn and discard the same pixels in the WMAP ILC map.

In Fig.~\ref{fig7} we plot the observed CCF for these six different
quasar subsamples, together with the theoretical CCF predictions of
WMAP5 best fit cosmological model. There is good agreement between the
theoretical and observed CCFs.

The values of CCF of the ``UVX-All'' subsample are consistent with
previous works \cite{gianna06,gianna08}. We find $0.22\pm 0.10\mu$K
and $0.26\pm 0.09\mu$K for the ``UVX-All'' and ``All-All''
subsamples at 1 deg, respectively. The median value of ``All-All''
subsample is larger than one of ``UVX-All'', due to the larger bias
and stellar contamination. The larger number of quasars in
``All-All'' subsample allow to slightly shrink the error bars. In
particular, we find that the high redshift subsample $z>1.5$ also
gives a non-vanishing cross-correlation signal, which is consistent
with the results of Ref.~\cite{hoetal08}. One of the possible
reasons could be that the high redshift subsample has a larger bias
and stellar contamination. Another speculative possibility is also
that a larger value for the ISW CCF than that implied by
$\Lambda$CDM could reflect a different underlying cosmological
model, characterized by an early departure from matter dominance and
onset of acceleration, as it happens in early dark energy and
modified gravity scenarios.

\section{Method and Datasets}
\label{method}

\subsection{Parametrization of Early Dark Energy}
\label{ede}

Due to the possible non-zero ISW signal at high redshift provided by
the early dark energy model, we will also consider here the EDE
mocker model introduced in Ref.~\cite{Mocker}:
\begin{equation}
w_{\rm EDE}(a)=-1+\left[1-\frac{w_0}{1+w_0}a^C\right]^{-1}~,
\end{equation}
where $a=1/(1+z)$ is the scale factor, $w_0$ is the present
equation-of-state of dark energy and $C$ characterizes the
``running" of the equation of state. Consequently, the evolution of
dark energy density can easily be obtained via energy conservation
as:
\begin{equation}
\frac{\rho_{\rm EDE}(a)}{\rho_{\rm
EDE}(1)}=\left[(1+w_0)a^{-C}-w_0\right]^{3/C}~.
\end{equation}

\begin{figure}[t]
\begin{center}
\includegraphics[scale=0.45]{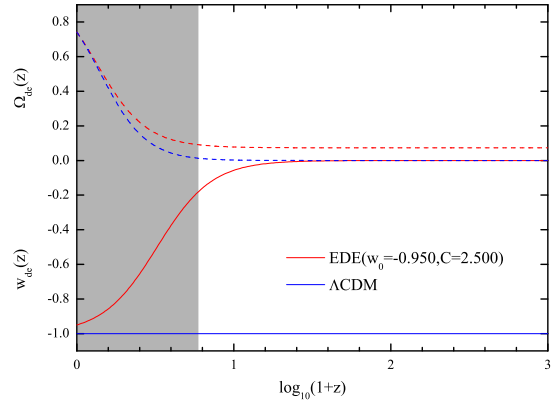}
\caption{The evolution of dark energy density and equation of state
for two models, $w_0=-0.95$ and $C=2.5$ (EDE) and pure $\Lambda$CDM
model. The shaded area represents the redshifts that is probed by
SDSS DR6-QSOs ($z\sim0-5$).\label{fig8}}
\end{center}
\end{figure}

In Fig.~\ref{fig8} we plot the dark energy density (upper part of
the panel) and equation of state (bottom part of the panel) as a
function of redshift for two different models: pure $\Lambda$CDM
(blue lines) and EDE (red lines) which has $(w_0,C)=(-0.95,2.50)$.
All these models fit the CMB and the lower redshift SNIa constraints
very well \cite{xiaviel09}. In the panel we also show as a shaded
vertical band the region in the redshift range $z=0-5$. This is the
redshift range we will be (mainly) focussing on in the rest of the
paper.

We stress that this is just one of the possible parameterizations
for early dark energy models, another one has been suggested by
Ref.~\cite{wetterich04} and recently used in Ref.~\cite{grossi08}.
However, we prefer to use the mocker model in order to compare with
the findings of Ref.~\cite{Mocker} and because this parametrization
has a smooth redshift derivative at low $z$ for $w(z)$. We note that
one of the most important parameters is the amount of dark energy
during the structure formation period and this is given by ($a_{\rm
eq}$ is the matter-radiation equality scale factor):
\begin{equation}
{\Omega}_{\rm EDE,sf} = -(\ln\, a_{\rm eq})^{-1}\int^{0}_{\ln a_{\rm
eq}} \Omega_{\rm EDE}\left(a\right){\rm d} \ln a~,
\label{eq:Omegasf}
\end{equation}
and we will also quote this value in the rest of the paper in order
to compare with other works and constraints as well (e.g.
Refs.~\cite{doran01,fedeli07}).

\begin{figure}[t]
\begin{center}
\includegraphics[scale=0.45]{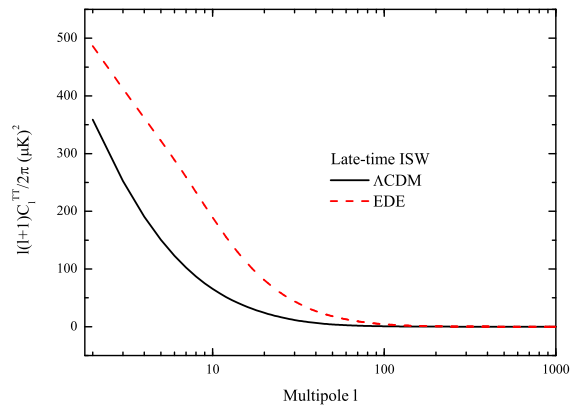}
\caption{The late-time ISW temperature power spectrum in two
different dark energy models: $w_0=-0.95$ and $C=2.5$ (EDE) and pure
$\Lambda$CDM model.\label{fig9}}
\end{center}
\end{figure}

\begin{figure*}[t]
\begin{center}
\includegraphics[scale=0.28]{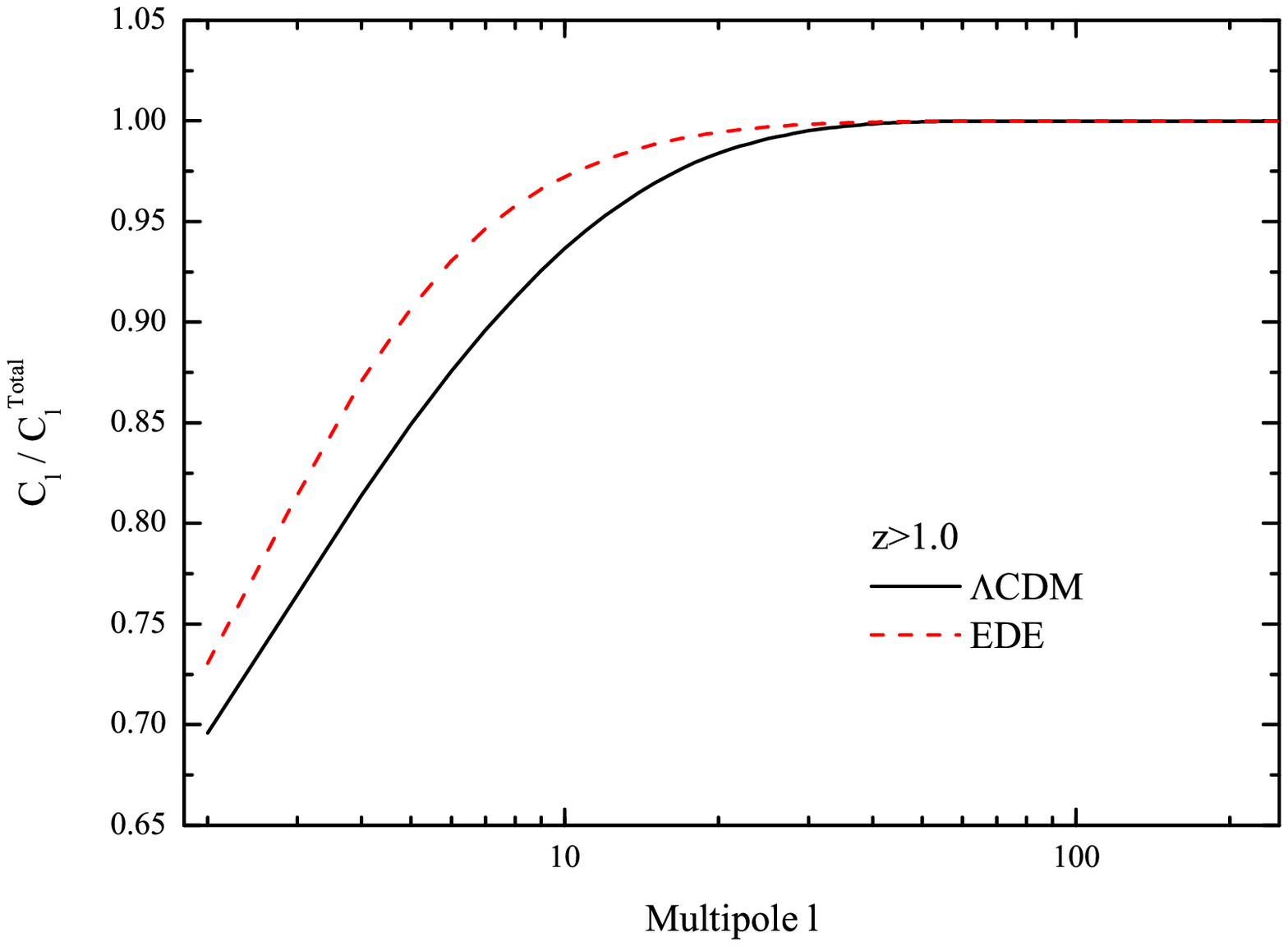}
\includegraphics[scale=0.28]{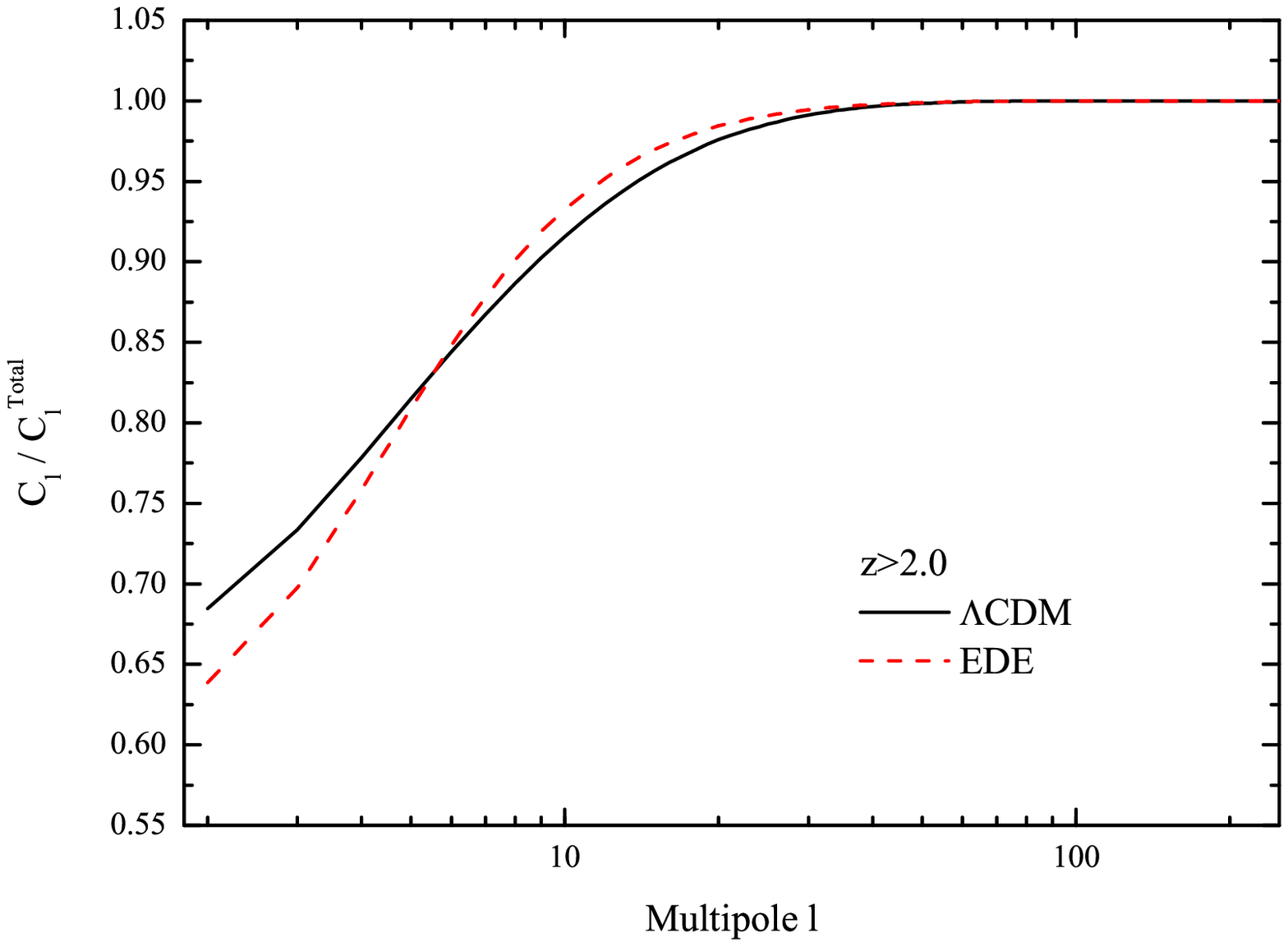}
\includegraphics[scale=0.28]{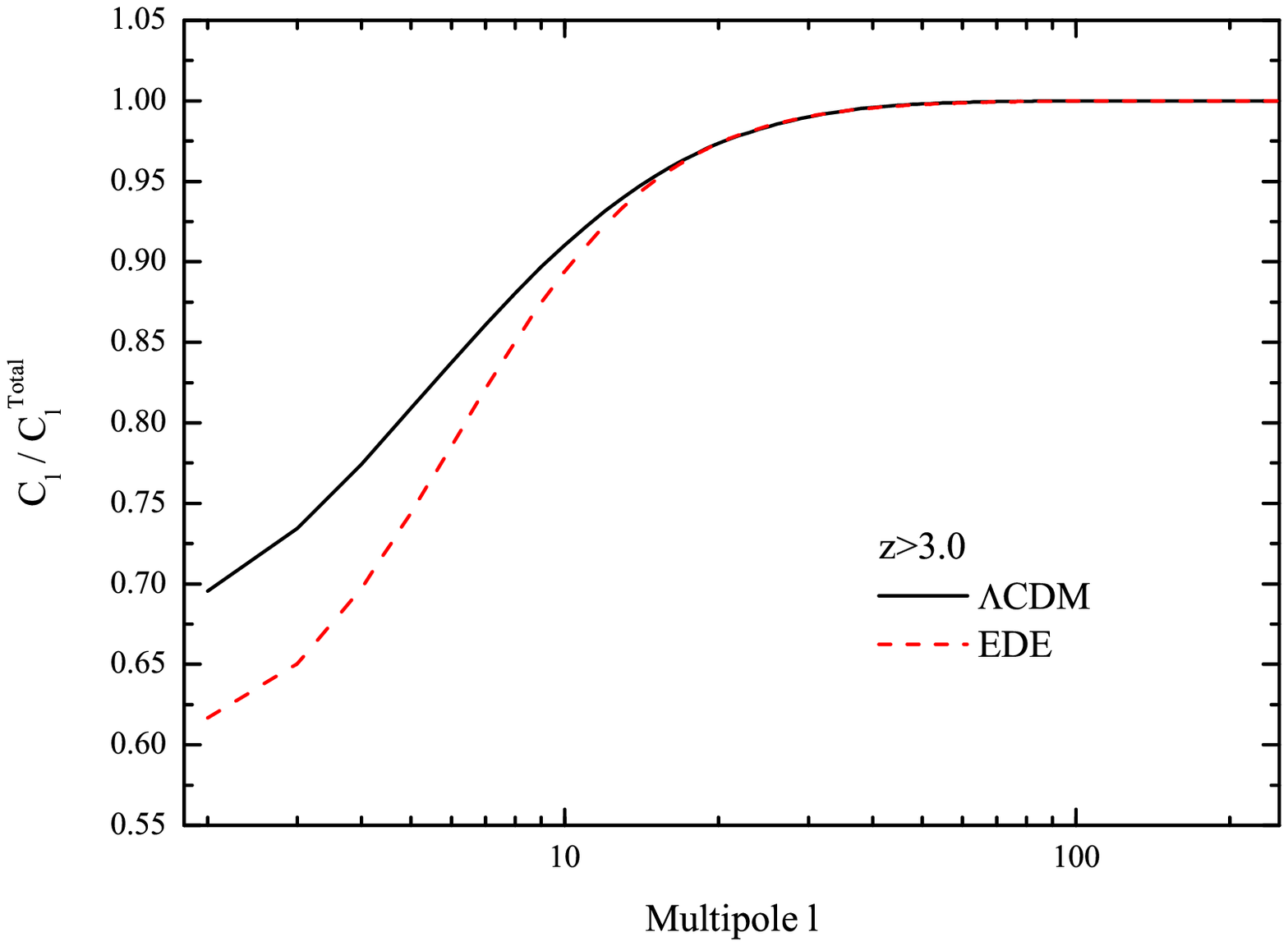}
\includegraphics[scale=0.28]{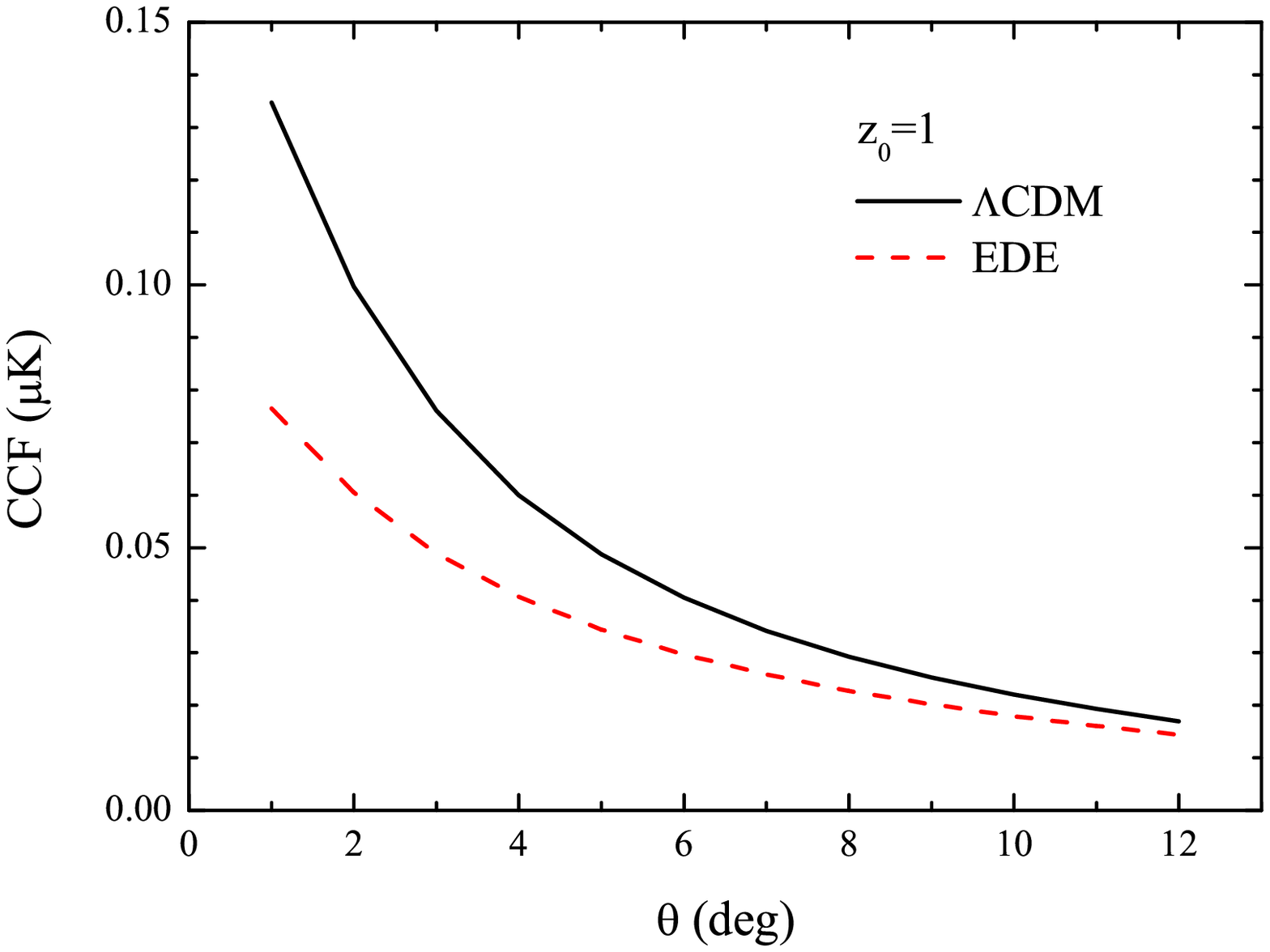}
\includegraphics[scale=0.28]{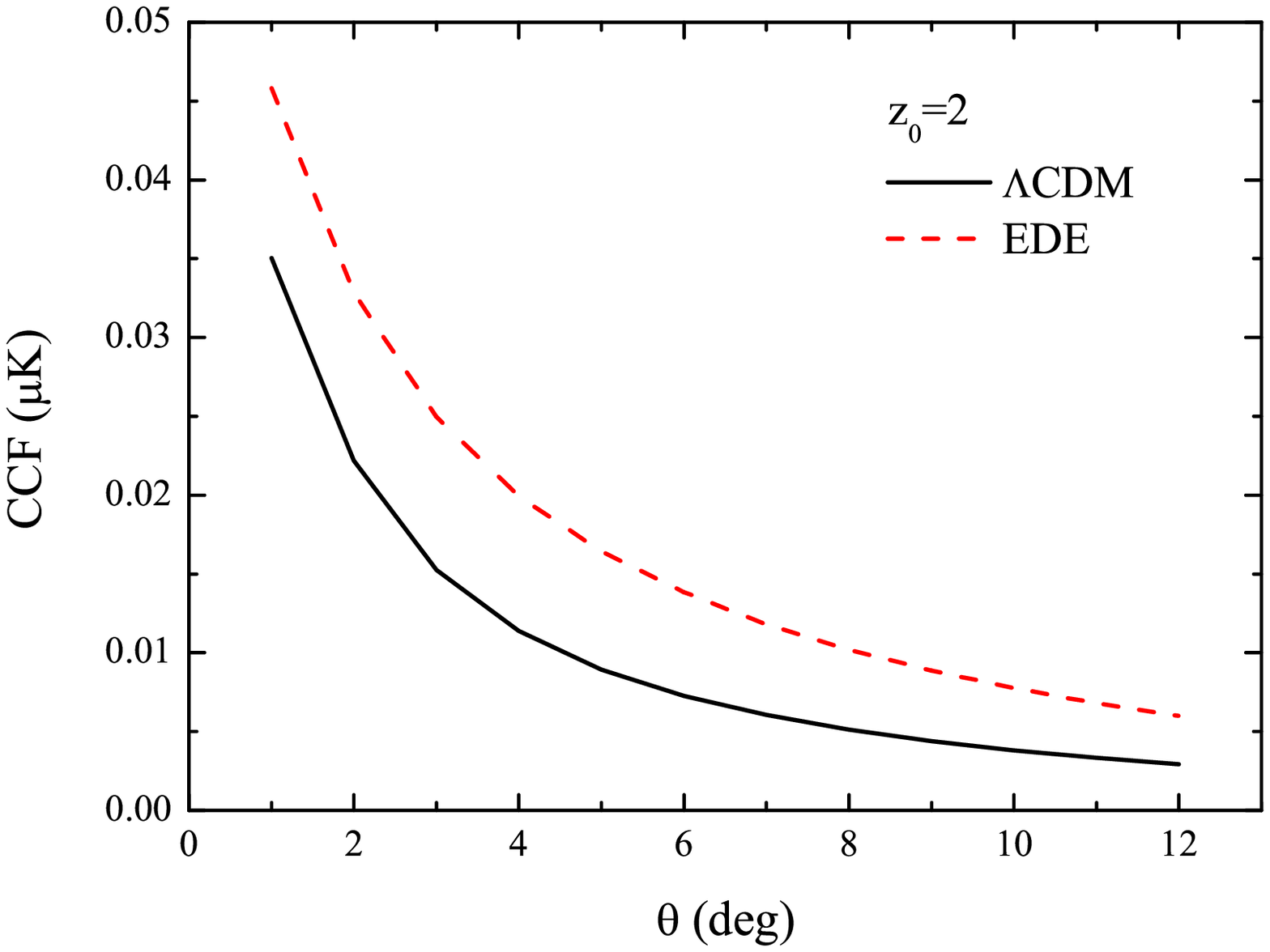}
\includegraphics[scale=0.28]{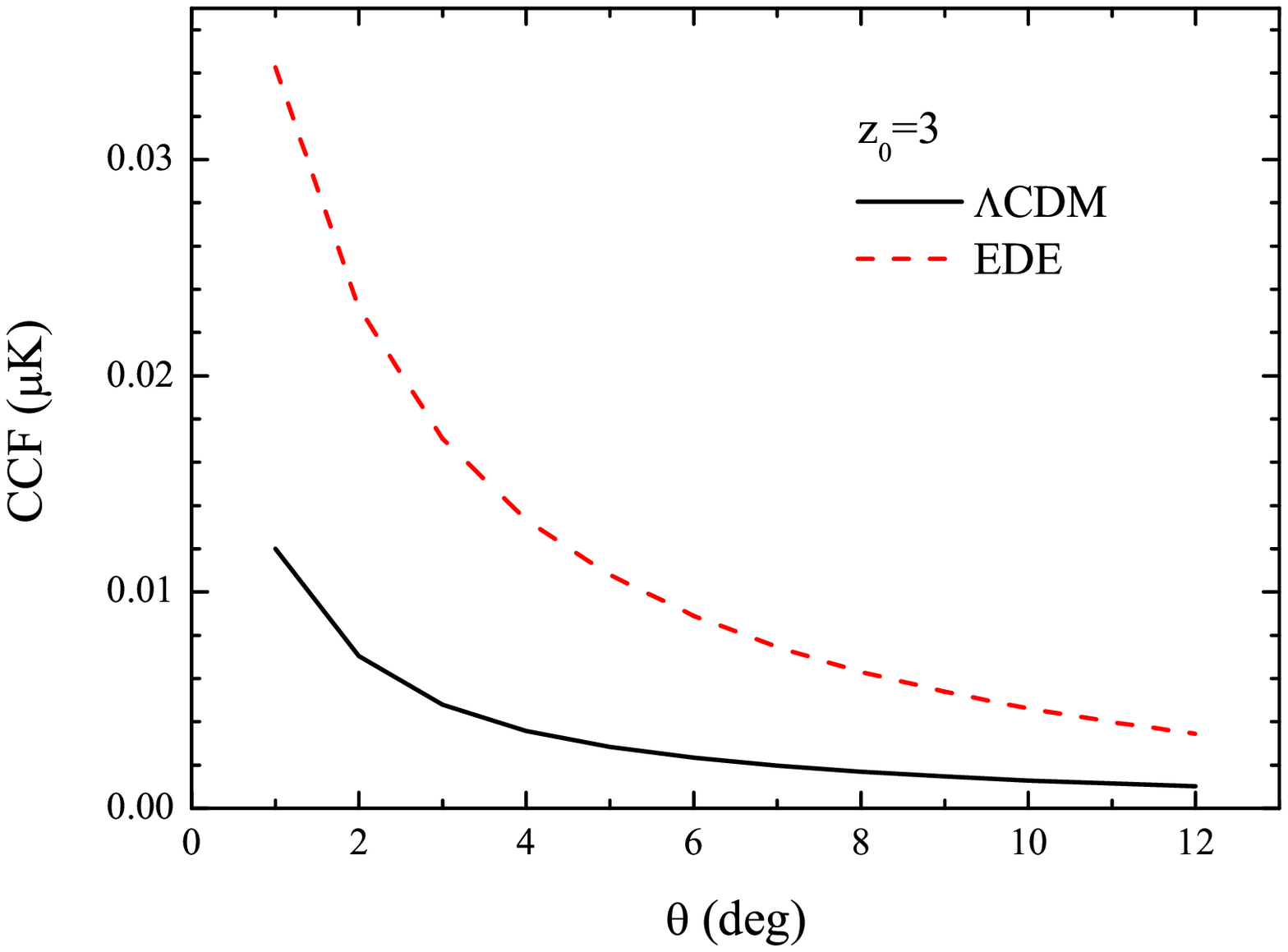}
\caption{Top: Ratio of the CMB temperature power spectrum with and
without the ISW effect at different low redshifts in two different
dark energy models: EDE (red dashed lines) and $\Lambda$CDM (black
solid lines). Bottom: The cross-correlation functions
$C^{gT}(\theta)$ between EDE Mocker model (red dashed lines) and
pure $\Lambda$CDM model (black solid lines). The redshift
distribution $dN/dz$ is assumed to be a Gaussian function centered
at three different mean redshifts $z_0=1,2,3$ comparison with a
$\sigma=0.5$. We fix bias to $b=2.2$ for illustrative purpose.
\label{fig10}}
\end{center}
\end{figure*}

In Fig.~\ref{fig9} we plot the late-time ISW temperature power
spectrum in the two different dark energy models analyzed. We find
that the early dark energy model produces a larger contribution to
the CMB primary power spectrum at low multipoles than the
$\Lambda$CDM model due to the different low redshift evolution:
differences up to a factor $\sim 2$ are present for $l$ values
smaller than $10$, reflecting the fact that ISW cross-correlation is
in place earlier than in $\Lambda$CDM, due to the early departure
from matter dominated expansion rate.

We also focus on the ISW effect at different redshifts investigating
EDE models. We modified the {\tt CAMB} code \cite{camb} to calculate
the contributions of ISW effect to the power spectrum prior to some
redshifts $z_\ast$, and neglect the ISW effect at low redshifts:
\begin{equation}
{\frac{\Delta T}{T}}^{\rm ISW}(\hat{\bf n})=0~,~{\rm
when}~z<z_\ast~.
\end{equation}
In Fig.~\ref{fig10} we plot the ratio of the CMB temperature power
spectra with and without the ISW effect at different low redshifts
and in two different dark energy models. We find that at the lowest
redshift $z<1$, the ISW contribution of $\Lambda$CDM is larger than
the EDE model. However, when the redshift becomes larger, the EDE
model contributes more to the ISW effect and to the CMB temperature
power spectrum. Therefore, when calculating the cross-correlation
power spectra between ISW and QSO survey with different mean
redshifts, we can see that the higher the mean redshift of QSO
survey is, the larger the cross-correlation $C_l^{\rm qT}$ becomes,
which is shown in Fig.~\ref{fig10}. From these plots it is clear how
the cross-correlation signal between QSO number density and CMB
temperature differs in the two models up to a factor three (the
differences at higher redshifts $z>3$ becoming even larger than
those shown here). Moreover, the redshift evolution of the ISW
signal is quite different in the two models and this is promising
for future studies that aim at investigating high-redshift
deviations from the standard model.

\subsection{ISW Likelihood Function}
\label{chi2}

In Refs.~\cite{gianna06,gianna08}, the authors firstly used the QSO
ACF data only to determine the bias $b$, when fixing the other
parameters and then they used the CMB-QSO CCF data to constrain the
significance of ISW signal or other cosmological parameters.
However, in order to make the whole analysis consistent, in our
calculations we use both the ACF and CCF data to constrain all the
parameters, including the bias $b$, efficiency $a$, ISW amplitude
$A_{\rm amp}$, which is defined in Eq.~(\ref{ampeq}), and other
cosmological parameters, simultaneously.

We then compare the theoretical ACF $c^{\rm qq}(\theta)$ and CCF
$c^{\rm qT}(\theta)$ with the observed values of ACF $\hat{c}^{\rm
qq}(\theta)$ and CCF $\hat{c}^{\rm qT}(\theta)$, respectively,
through the Gaussian likelihood function:
\begin{eqnarray}
\mathcal{L}_{\rm ACF}&=&(2\pi)^{-N/2}[{\rm
det}C_{ij}]^{-1/2}\nonumber\\
&\times&\exp\left[-\sum_{i,j}\frac{C^{-1}_{ij}(\hat{c}^{\rm
      qq}_i-{c}^{\rm qq}_i)(\hat{c}^{\rm qq}_j-{c}^{\rm qq}_j)}{2}\right]~,\\
\mathcal{L}_{\rm CCF}&=&(2\pi)^{-N/2}[{\rm
det}C'_{ij}]^{-1/2}\nonumber\\
&\times&\exp\left[-\sum_{i,j}\frac{C'^{-1}_{ij}(\hat{c}^{\rm
      qT}_i-{c}^{\rm qT}_i)(\hat{c}^{\rm qT}_j-{c}^{\rm qT}_j)}{2}\right]~,
\end{eqnarray}
where $C_{ij}$ and $C'_{ij}$ are the DR6-QSO auto-correlation
function and CMB-QSO cross-correlation function covariance matrix.

\subsection{Other Current Datasets}
\label{current}

Besides the ACF and CCF data, we will rely here on the following
cosmological probes: ${\rm i})$ CMB anisotropies and polarization;
${\rm ii})$ baryonic acoustic oscillations in the galaxy power
spectra; ${\rm iii})$ SNIa distance moduli.

In the computation of CMB power spectra we have included the WMAP
five-year (WMAP5) temperature and polarization power spectra with
the routines for computing the likelihood supplied by the WMAP team
\cite{Komatsu:2008hk,hinshaw09,gold09,WMAP5:Other1,WMAP5:Other2,WMAP5:Other3}.

BAOs (Baryonic Acoustic Oscillations) have been detected in the
current galaxy redshift survey data from the SDSS and the Two-degree
Field Galaxy Redshift Survey (2dFGRS)
\cite{Eisenstein:2005su,Cole:2005sx,Huetsi:2005tp,BAO}. The BAO can
directly measure not only the angular diameter distance, $D_A(z)$,
but also the expansion rate of the universe, $H(z)$, which is
powerful for studying dark energy \cite{Albrecht:2006um}. Since
current BAO data are not accurate enough for extracting the
information of $D_A(z)$ and $H(z)$ separately \cite{Okumura:2007br},
one can only determine an effective distance
\cite{Eisenstein:2005su}:
\begin{equation}
D_v(z)\equiv\left[(1+z)^2D_A^2(z)\frac{cz}{H(z)}\right]^{1/3}~.
\end{equation}
In this paper we use the Gaussian priors on the distance ratios
$r_s(z_d)/D_v(z)$:
\begin{eqnarray}
r_s(z_d)/D_v(z=0.20)&=&0.1980\pm0.0058~,\nonumber\\
r_s(z_d)/D_v(z=0.35)&=&0.1094\pm0.0033~,
\end{eqnarray}
with a correlation coefficient of $0.39$, extracted from the SDSS
and 2dFGRS surveys \cite{BAO}, where $r_s$ is the comoving sound
horizon size and $z_d$ is drag epoch at which baryons were released
from photons given by Ref.~\cite{Eisenstein:1997ik}.

The SNIa data provide the luminosity distance as a function of
redshift which is also a very powerful measurement of dark energy
evolution. The supernovae data we use in this paper are the recently
released Union compilation (307 samples) from the Supernova
Cosmology project \cite{Kowalski:2008ez}, which include the recent
samples of SNIa from the (Supernovae Legacy Survey) SNLS and ESSENCE
survey, as well as some older data sets, and span the redshift range
$0\lesssim{z}\lesssim1.55$. In the calculation of the likelihood
from SNIa we have marginalized over the nuisance parameter as done
in Ref.~\cite{SNMethod}.

Furthermore, we make use of the Hubble Space Telescope (HST)
measurement of the Hubble parameter $H_{0}\equiv
100\,h$~km~s$^{-1}$~Mpc$^{-1}$ by a Gaussian likelihood function
centered around $h=0.72$ and with a standard deviation $\sigma=0.08$
\cite{HST}.

\subsection{Future Datasets}
\label{future}

In order to forecast future measurements we will use the same
observables as before without BAO.

For the simulation with Planck \cite{Planck}, we follow the method
given in Ref.~\cite{xiaplanck} and mock the CMB temperature (TT) and
polarization (EE) power spectra and temperature-polarization
cross-correlation (TE) by assuming a given fiducial cosmological
model. In Table I, we list the assumed experimental specifications
of the future (mock) Planck measurement.

\begin{table}
TABLE I. Assumed experimental specifications for the mock
Planck-like measurements. The noise parameters $\Delta_T$ and
$\Delta_P$ are given in units of $\mu$K-arcmin.
\begin{center}
\begin{tabular}{cccccc}
\hline \hline

$f_{\rm sky}$~ & ~$l_{\rm max}$~ & (GHz) &
~$\theta_{\rm fwhm}$~ & ~$\Delta_T$~~ & ~~$\Delta_P$~ \\

\hline

 0.65 & 2500 & 100 & 9.5' & 6.8 & 10.9 \\
      &      & 143 & 7.1' & 6.0 & 11.4 \\
      &      & 217 & 5.0' & 13.1 & 26.7 \\

\hline \hline
\end{tabular}
\end{center}
\end{table}

The proposed satellite SNAP\footnote{Available at
http://snap.lbl.gov/.} (Supernova / Acceleration Probe) will be a
space based telescope with a one square degree field of view that
will survey the whole sky. It aims at increasing the discovery rate
of SNIa to about $2000$ per year in the redshift range $0.2<z<1.7$.
In this paper we simulate about $2000$ SNIa according to the
forecast distribution of the SNAP \cite{Kim:2003mq}. For the error,
we follow the Ref.~\cite{Kim:2003mq} which takes the magnitude
dispersion to be $0.15$ and the systematic error $\sigma_{\rm
sys}=0.02\times z/1.7$. The whole error for each data is given by
$\sigma_{\rm mag}(z_i)=\sqrt{\sigma^2_{\rm
sys}(z_i)+0.15^2/{n_i}}~$, where $n_i$ is the number of supernovae
of the $i'$th redshift bin.

For the future ISW ACF and CCF data, we simulate two mock datasets
from the best fit values of data combination
WMAP5+BAO+SNIa+``UVX-All'' in the $\Lambda$CDM and EDE models,
respectively. We also set $b=2.0$, $a=97\%$ and $A_{\rm amp}=1$ with
error bars on these parameters that are reduced by a factor three.
We use directly the covariance matrix taken from present data and
divide it by a factor nine. This improvement could be achievable by
next generation of large-scale surveys such as the Large Synoptic
Survey Telescope (LSST, \cite{Tyson02}), the Panoramic Survey
Telescope and Rapid Response System (Pan-STARRS, \cite{Kaiser02})
and the Dark Energy Survey (DES; The Dark Energy Survey
Collaboration 2005) are likely to allow for an order of magnitude
improvement in the number of QSOs which will account for the factor
three considered here. Also, the coverage of the sky fraction is at
present of the order of $20\%$ by SDSS DR6, so an all-sky survey
will already give a factor two improvement (we assume that the error
bars scale as $\sqrt{N_{\rm QSOs}}$, which is reasonable if these
are independent). Another possible improvement in this direction is
the use of type-2 QSOs instead of the type-1 used here that could
further decrease the error bars of the sample (see the discussion in
\cite{richards09}).

\section{Numerical Results}
\label{results}

In our analysis, we perform a global fitting using the {\tt CosmoMC}
package \cite{cosmomc} a Monte Carlo Markov Chain (MCMC) code, which
has been modified to calculate the theoretical ACF and CCF. We assume
purely adiabatic initial conditions and a flat universe, with no
tensor contribution. We vary the following cosmological parameters
with top-hat priors: the dark matter energy density $\Omega_{\rm c} h^2
\in [0.01,0.99]$, the baryon energy density $\Omega_{\rm b} h^2 \in
[0.005,0.1]$, the primordial spectral index $n_{\rm s} \in [0.5,1.5]$, the
primordial amplitude $\log[10^{10} A_{\rm s}] \in [2.7,4.0]$ and the angular
diameter of the sound horizon at last scattering $\theta \in
[0.5,10]$. For the pivot scale we set $k_{\rm s0}=0.05\,$Mpc$^{-1}$. When
CMB data are included, we also vary the optical depth to reionization
$\tau \in [0.01,0.8]$. We do not consider any massive neutrino
contribution.  From the parameters above the MCMC code derives the
reduced Hubble parameter $H_0$, the present matter fraction
$\Omega_{{\rm m}0}$, $\sigma_8$, and $\Omega_{\rm EDE,sf}$, so, these
parameters have non-flat priors and the corresponding bounds must be
interpreted with some care. There are three more parameters related to
the ACF and CCF data: the constant bias $b$, the efficiency of quasar
catalog $a$ and the ISW amplitude $A_{\rm amp}$ which is defined by:
\begin{equation}
\bar{C}^{\rm qT}(\theta)=A_{\rm amp}C^{\rm qT}(\theta)~,\label{ampeq}
\end{equation}
where $\bar{C}^{\rm qT}$ and $C^{\rm qT}$ are the observed and theoretical
CCF. In addition, {\tt CosmoMC} imposes a weak prior on the Hubble
parameter: $h \in [0.4,1.0]$. Furthermore, we also include the
perturbations of dynamical dark energy models generally as done in
Refs.~\cite{pert,zhaopert05,xiapert06}.

\begin{figure}[t]
\begin{center}
\includegraphics[scale=0.4]{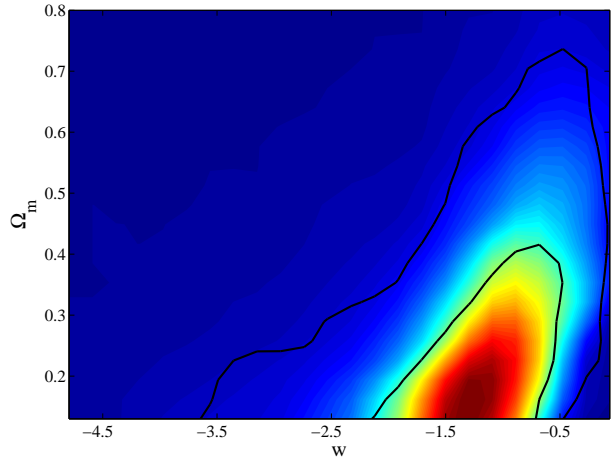}
\caption{Two-dimensional (marginalized) contours of $\Omega_{\rm
m0}$ and constant $w_{\rm DE}$ from the ISW data only.\label{fig11}}
\end{center}
\end{figure}

\subsection{ISW Only}
\label{iswonly}

Firstly, we use the ACF and CCF data of ``UVX-All'' subsample only
to constrain the dark energy model with a constant equation of state
$w_{\rm DE}$. Because at present the ISW ACF and CCF data have
relatively large error bars that do not allow to use this
measurement in a very competitive way compared to other cosmological
probes, in our calculation we have fixed the other cosmological
parameters to the WMAP5 best fit values, as well as the bias,
efficiency and amplitude parameters. In Fig.~\ref{fig11} we plot the
two-dimensional constraints on ($w_{\rm DE}$,$\,\Omega_{\rm m0}$).
The results show that the ISW data only could give very weak
constraints on the background parameters: $w_{\rm
DE}=-1.24\pm0.62~(1\,\sigma)$ and the $2\,\sigma$ upper limit of
$\Omega_{\rm m0}<0.55$, which is consistent with the pure
$\Lambda$CDM model. This result is also consistent with some
previous works
\cite{Corasaniti05,Pietrobon:2006gh,gianna06,gianna08}.

\begin{table}
TABLE II. The results of quasar bias and efficiency of quasar
catalog in two dark energy models.
\begin{center}
\begin{tabular}{lccc}
\hline\hline
Subsample~~&Mean Redshift $\bar{z}$&~~~Bias $b$~~~~&~~Efficiency $a$\\
\hline \multicolumn{4}{c}{$\Lambda$CDM model}\\
All-All &$1.80$ & $3.79\pm0.34$  & ~~$88.1\%\pm0.5\%$\\
All-Low &$0.90$& $1.46\pm0.29$  & ~~$97.1\%\pm0.5\%$\\
All-High &$2.28$& $5.01\pm0.81$  & ~~$80.9\%\pm0.7\%$\\
\hline
UVX-All &$1.49$& $2.18\pm0.22$  & ~~$96.8\%\pm0.5\%$\\
UVX-Low &$0.90$& $0.92\pm0.30$  & ~~$97.5\%\pm0.8\%$\\
UVX-High &$2.02$& $2.87\pm0.39$  & ~~$95.1\%\pm0.6\%$\\
\hline\hline \multicolumn{4}{c}{EDE Mocker model}\\
UVX-All &$1.49$& $2.33\pm0.31$  & ~~$96.8\%\pm0.5\%$\\
UVX-Low &$0.90$& $1.05\pm0.33$  & ~~$97.5\%\pm0.8\%$\\
UVX-High &$2.02$& $2.98\pm0.43$  & ~~$95.1\%\pm0.6\%$\\
\hline\hline
\end{tabular}
\end{center}
\end{table}

\subsection{Quasar Bias and Efficiency}
\label{bias}

In our calculations we simply assume that the quasar bias is constant in
the redshift region considered. Table II shows the results of quasar
bias and efficiency of six subsamples in two dark energy models.

For the ``UVX-All'' subsample, we obtain a value for the bias
$b=2.18\pm0.22$ ($1\,\sigma$) in the $\Lambda$CDM framework, also
consistent with previous measurements
\cite{gianna06,myersetal06,croom04,myersetal07}. However, we find
that this bias result depends  on the dark energy model we choose.
In the EDE Mocker model, the bias becomes $b=2.33\pm0.31$
($1\,\sigma$). Thus, the mean value of quasar bias becomes slightly
higher in an EDE cosmology by $\sim 10\%$. This effect is generally
present in the subsamples at different redshifts.

On the other hand, we find that the efficiency of ``UVX-All''
subsample is $a=96.8\pm0.5\%$, also consistent with
Ref.~\cite{richards09}. This efficiency does not depend on the
different dark energy model, which is reasonable since this
 is an intrinsic feature of the quasar catalog.

In the high redshift subsample, the quasar bias will also become
higher and the efficiency becomes clearly smaller. As we mentioned
before, at high redshift the stellar contamination will become
large. The cross term $\epsilon(\theta)$ will be comparable to the
star ACF $\hat{c}^{ss}(\theta)$ in Eq.~(\ref{crossterm}). In order to
obtain the correct efficiency, we should consider this cross terms
in our calculations. If we neglect this cross term, the efficiency
will be incorrectly suppressed, since in this case one should need a
larger stellar contamination to contribute more ACF. For example,
when using ``All-All'' subsample to do the calculations, if we
neglect the cross terms $\epsilon(\theta)$, the efficiency is
$82.5\%$, which is consistent with Ref.~\cite{richards09}. However,
if taking $\epsilon(\theta)$ into account, the efficiency rises
to $88.1\%$.

\begin{table}
TABLE III. Estimates of the quasar bias, from five photometric
redshift bins.
\begin{center}
\begin{tabular}{ccc}
\hline \hline

Redshift Bins & Mean Redshift $\bar{z}$ & Bias $b$\\

\hline

$0.75<z<1.25$ & $1.02$ & $1.58\pm0.39$ \\
$1.25<z<1.55$ & $1.43$ & $1.85\pm0.78$ \\
$1.55<z<1.95$ & $1.73$ & $2.23\pm0.87$ \\
$1.95<z<2.20$ & $2.06$ & $3.28\pm0.86$ \\
$2.20<z<4.00$ & $2.46$ & $5.29\pm1.31$ \\

\hline \hline
\end{tabular}
\end{center}
\end{table}

From the above analysis, we find that the quasar bias could evolve
with redshift. Models in which the bias varies with redshift have
been proposed and investigated (e.g. Refs.
\cite{matarrese97,moscardini98}). Here, we also give the constraint
on the evolution of the QSO bias with redshift. In our calculation
we are not using more sophisticated models as those in
Refs.~\cite{moscardini98,martini01,porciani04}. We only split the
whole quasar ``UVX-All'' subsample into five photometric redshift
bins: $[0.75,1.25]$, $[1.25,1.55]$, $[1.55,1.95]$, $[1.95,2,2]$ and
$[2.2,4.0]$, containing $\sim10^5$ objects each, and assume a
constant bias in each of them. Then we calculate their auto
correlation function to determine their quasar bias. In this
analysis we do not include data at $z<0.75$, since in this low
redshift the subsample is highly contaminated by the galaxies
\cite{myersetal06,weinstein04}.

\begin{figure}[t]
\begin{center}
\includegraphics[scale=0.43]{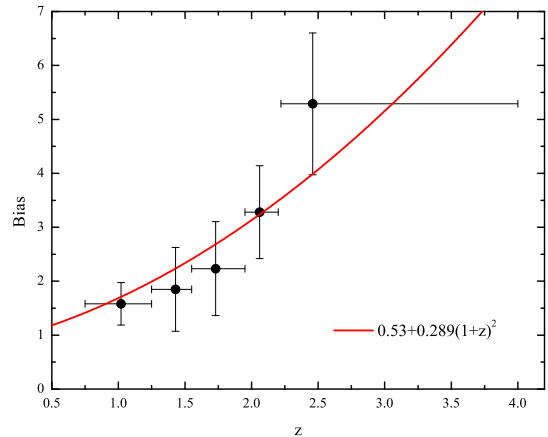}
\caption{Estimates of the quasar bias and their $1\,\sigma$ error
bars from five photometric redshift bins. The red solid line
$b(z)=0.53+0.289(1+z)^2$ is the empirical fit derived by
Ref.~\cite{croom04}.\label{fig12}}
\end{center}
\end{figure}

In Table III and Fig.~\ref{fig12} we show the estimates of the
quasar bias in these five photometric redshift bins. We find that
our results are in overall agreement with previous works
\cite{myersetal06,croom05,shen07}. In Fig.~\ref{fig12} we also plot
the empirical fit derived by directly measuring the real-space
clustering of spectroscopically confirmed QSOs in
Ref.~\cite{croom04}. Our results are consistent with this empirical
fit very well. Furthermore, this analysis also allow to estimate
important quantities such as the mass of the host haloes, QSO duty
cycle and the Mass-Luminosity relation (see Ref.~\cite{martini01}).

\subsection{Amplitude of ISW Signal}
\label{amp}

In previous works \cite{gianna08,hoetal08}, cosmological parameters
were fixed and the amplitude of ISW signal was calculated from the
cross-correlation data only. However, generally speaking this
amplitude parameter is highly dependent on the cosmological
parameters we use. Currently, the constraints on the cosmological
parameters are not affected significantly by fixing the three
parameters; however, this is because the present data are not very
constraining. On the other hand, in our calculations we find that
the constraints on bias and amplitude are different between LCDM and
EDE models (see Table II and IV). This fact demonstrates that
keeping these parameters fixed will lead to biased results. In other
words, had we used the bias and amplitude obtained from the LCDM
model to constrain the EDE model, we would have obtained different,
biased results. Therefore, here we constrain all these parameters at
the same time using the MCMC.

The results are shown in Table IV. The significance in sigmas is
obtained by the simple calculation, $S/N=A/\sigma_A$. For the
``UVX-All'' subsample the significance of ISW is about
$2.7\,\sigma$, which is consistent with other works
\cite{gianna06,gianna08}. Furthermore, we also find that the low
redshift subsample also gives about $2.3\,\sigma$ significance,
which is reduced as compared to the ``All'' sample due the smaller
quasar number of the low redshift subsample.

\begin{table}
TABLE IV. The results of amplitude of ISW signal and their
significance in two dark energy models.
\begin{center}
\begin{tabular}{lcc}
\hline\hline
Subsample~~&~~Amplitude $A_{\rm amp}$~~&~~S/N~~~~\\
\hline \multicolumn{3}{c}{$\Lambda$CDM Model}\\
All-All &$1.51\pm0.73$&$2.1\,\sigma$\\
All-Low &$3.05\pm1.31$&$2.3\,\sigma$\\
All-High &$2.32\pm1.53$&$1.5\,\sigma$\\
\hline
UVX-All &$2.06\pm0.75$&$2.7\,\sigma$\\
UVX-Low &$4.36\pm1.82$&$2.4\,\sigma$\\
UVX-High &$2.49\pm1.57$&$1.6\,\sigma$\\
\hline\hline \multicolumn{3}{c}{EDE Mocker Model}\\
UVX-All &$1.91\pm0.75$&$2.5\,\sigma$\\
UVX-Low &$3.96\pm1.78$&$2.2\,\sigma$\\
UVX-High &$2.22\pm1.47$&$1.5\,\sigma$\\
\hline\hline
\end{tabular}
\end{center}
\end{table}

The most interesting result is that the high redshift $z>1.5$
subsample reports a signal of ISW effect at $\sim1.5\,\sigma$
confidence level. As we know, in the pure $\Lambda$CDM model with
the bias $b\sim1$, the prediction of ISW effect should be close to
zero in the high redshift $A_{\rm amp}\sim0$. However, in our
analysis we obtain a higher mean value and a smaller error bar.
Consequently, the amplitude of ISW signal is larger than a
null-detection at about $\sim1.5\,\sigma$ confidence level. As we
discussed before, this interesting non-vanishing ISW signal in the
high redshift subsample may be due to the larger bias and stellar
contamination at this high redshift, or some different underlying
cosmological models reflecting an early departure for matter
dominance in the expansion rate. The large number of quasar in the
sample is also helpful in shrinking the error bar and enhancing the
significance of the signal.  Another intriguing result that we can
appreciate from Table IV is that the discrepancies with the value
$A_{\rm amp}=1$, which labels the perfect agreement between theory
and data, are more severe for the lower redshift ranges than for
those at high redshift. All the high-redshift samples are in fact in
agreement at the $1\,\sigma$ level with the theoretical predictions
either for the EDE or $\Lambda$CDM cosmologies. As for the
low-redshift samples the discrepancies are at the $\sim 2\,\sigma$
level confirming the results of Ref.~\cite{hoetal08}.

As we show in Fig.~\ref{fig10}, at high redshift the QSO-temperature
cross-correlations of the EDE model will be larger than that
predicted in the $\Lambda$CDM framework. When combining the obtained
value for the bias, the EDE model should give larger CCF values than
those of $\Lambda$CDM model for  a given cosmology. Thus, in our
calculations we find that the amplitude of the ISW signal in EDE
model, inversely proportional to the theoretical CCF values (see
Eq.~(\ref{ampeq})), is slightly smaller than the one of $\Lambda$CDM
model.  Although other mechanisms could be in place to explain this
discrepancy, either involving not properly understood systematic
effects or large scale structures such as super-clusters or large
voids \cite{granett08}, a not negligible amount of dark energy at
high redshift could also help in reducing the statistical
significance of this result.

\subsection{Cosmological Constraints}
\label{lcdm}

\begin{table*}
TABLE V. Constraints on the $\Lambda$CDM and early dark energy model
from the current observations. Here we show the mean values and
$1\,\sigma$ error bars. For some parameters that are only weakly
constrained we quote the $95\%$ upper limit.
\begin{center}

\begin{tabular}{lccc}

\hline\hline

~~Parameter~~~~&~~~{WMAP5 Only}~~~~&~~~WMAP5+ISW~~~~&~~~All Datasets~~~~\\
\hline
\multicolumn{4}{c}{$\Lambda$CDM Model}\\

~~$\Omega_{\rm m}$ & $0.261\pm0.030$ & $0.261\pm0.028$ & $0.273\pm0.019$\\
~~$\sigma_8$ & $0.797\pm0.035$ & $0.795\pm0.032$ & $0.805\pm0.027$\\
~~$H_0$ & $71.4\pm2.6$ & $71.3\pm2.5$ & $70.2\pm1.7$\\
~~${\Omega}_{\rm{EDE,sf}}$&$0.0559\pm0.0050$&$0.0557\pm0.0048$&$0.0537\pm0.0032$\\
\hline\hline
\multicolumn{4}{c}{EDE Mocker Model}\\

~~$w_0$ & $<-0.694$ & $<-0.708$ & $<-0.909$\\
~~$C$ & $<2.950$ & $<2.623$ & $<3.214$\\
~~$\Omega_{\rm m}$ & $0.307\pm0.050$ & $0.303\pm0.048$ & $0.272\pm0.021$\\
~~$\sigma_8$ & $0.716\pm0.072$ & $0.714\pm0.070$ & $0.744\pm0.049$\\
~~$H_0$ & $66.1\pm4.4$ & $66.3\pm4.3$ & $69.3\pm1.8$\\
~~$\Omega_{\rm EDE}(z_{\rm lss})$ & $<0.037$ & $<0.036$ & $<0.026$\\
~~${\Omega}_{\rm{EDE,sf}}$&$0.0681\pm0.0144$&$0.0682\pm0.0139$&$0.0644\pm0.0104$\\
\hline\hline
\end{tabular}
\end{center}
\end{table*}

Finally we present  constraints on the cosmological parameters
from the ISW data, combining with the WMAP5, BAO and SNIa datasets,
in two dark energy models. Here we only use the ACF and CCF data
from the ``UVX-All'' subsample. The other five subsamples also
give similar results. In our calculations we do not follow the
previous works \cite{gianna06,gianna08} by fixing the other three
parameters, $b$, $a$ and $A_{\rm amp}$, to be their best fit values.
In Table V we show the constraints on some related cosmological
parameters from three different data combinations: WMAP5 only,
WMAP5+ISW, and All datasets. And we particularly pay attention to
the ISW contribution by comparing the results between WMAP5 and
WMAP5+ISW.

\begin{figure}[t]
\begin{center}
\includegraphics[scale=0.41]{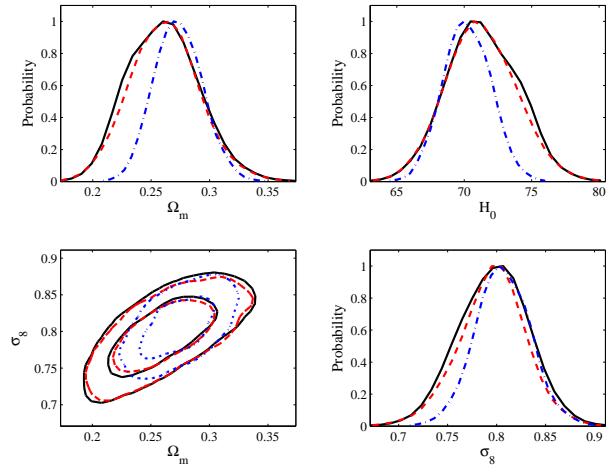}
\caption{Marginalized one-dimensional and two-dimensional likelihood
($1,2\sigma$ contours) constraints on the parameters $\Omega_{\rm m}$, $H_0$ and $\sigma_8$
in the $\Lambda$CDM model from different present data combinations:
WMAP5 only (black solid lines), WMAP5+ISW (red dashed lines) and All
datasets (blue dash-dot lines).\label{fig13}}
\end{center}
\end{figure}

\subsubsection{$\Lambda$CDM Model}

Firstly, we consider the pure $\Lambda$CDM model. In
Fig.~\ref{fig13} we show the one dimensional marginalized likelihood
distributions of some cosmological parameters from three data
combinations. From Table V we can find that the combined constraints
from WMAP5+ISW are only slightly improved over using WMAP5 only,
since at present constraints from the ISW data are still very weak
and in the calculations we only consider the quasar catalog and
neglect other low redshift tracers which could give powerful ISW
constraints \cite{gianna08,hoetal08,raccanellietal08}. We also show
the two dimensional contour in the ($\Omega_{\rm m}$,$\,\sigma_8$)
panel. When adding the ISW data, the constraint becomes slightly
more stringent.

When combining all the datasets together, the constraints tighten
significantly: the error bars of some parameters are reduced by a
factor of $1.5$. These is due to the constraining power of SNIa and
BAO.  These results are consistent with some previous works
\cite{Komatsu:2008hk,xiaglobal08} and also with the recent findings
based on the clustering of luminous red galaxies of the DR7
\cite{reid}. Thus, present data sets allow to constrain the amount of
dark energy in the structure formation era at the percent level.

\begin{figure}[t]
\begin{center}
\includegraphics[scale=0.41]{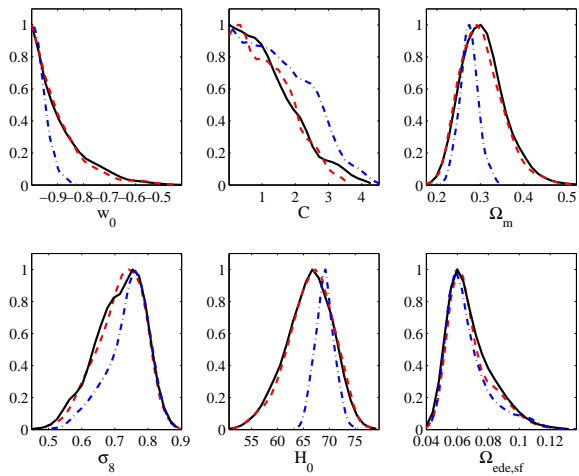}
\caption{One-dimensional marginalized likelihood constraints on the
dark energy parameters $w_0$ and $C$, as well as $\Omega_{\rm m}$,
$H_0$, $\sigma_8$ and $\Omega_{\rm EDE,sf}$ from different present
data combinations: WMAP5 only (black solid lines), WMAP5+ISW (red
dashed lines) and All datasets (blue dash-dot lines).\label{fig14}}
\end{center}
\end{figure}

\subsubsection{EDE Mocker Model}

Due to the lack of  cosmological probes, the behaviour of the
dark energy component is very poorly constrained in the redshift
range $2 < z < 1100$.  Therefore, when using WMAP5 data only, the
constraints on the parameters $w_0$ and $C$, describing the equation
of state of early dark energy, are very weak, namely the $95\%$
upper limits are $w_0<-0.694$ and $C<2.950$. Consequently, current
observations still allow very large amount of early dark energy at
high redshift $z \sim 1090$ as $\Omega_{\rm EDE}(z_{\rm lss}) <
0.037$ ($95\%$ C.L.), which is consistent with the results obtained
by Refs.~\cite{xiaviel09,Mocker,leeng02}. Early dark energy models
with a non-negligible fraction of dark energy density still fit the
CMB data very well.

\begin{figure}[t]
\begin{center}
\includegraphics[scale=0.38]{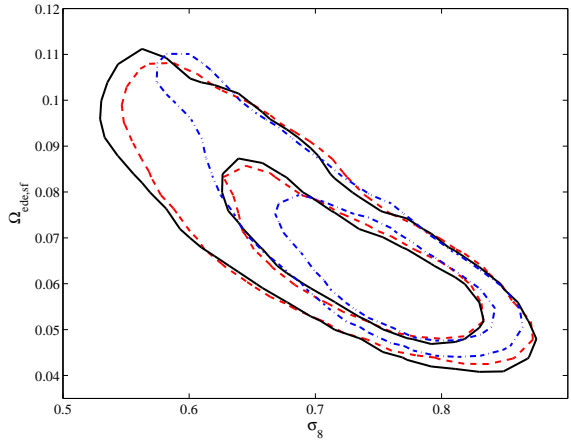}
\caption{Two dimensional contour in the ($\sigma_8$,$\,\Omega_{\rm
EDE,sf}$) panel from different present data combinations: WMAP5 only
(black solid lines), WMAP5+ISW (red dashed lines) and All datasets
(blue dash-dot lines).\label{fig15}}
\end{center}
\end{figure}

When comparing the results of $\Lambda$CDM and EDE model, the error
bars of some parameters are significantly enlarged by a factor of
two, shown in Fig.\ref{fig14}, due to the degeneracies between the
dark energy parameters and other background parameters. We find that
the constraint on the parameter $\Omega_{\rm EDE,sf}$ becomes rather
weak for EDE. The current constraint on $\Omega_{\rm EDE,sf}$ is
$\Omega_{\rm EDE,sf}=0.0681\pm0.0144$ at $1\,\sigma$ confidence
level, which is obviously higher than the pure $\Lambda$CDM model,
although the two agree at $1\sigma$ level: $\Omega_{\rm
EDE,sf}=0.0559\pm0.0050$ ($1\,\sigma$). This is because of the
higher dark energy abundance at high redshifts with respect to
$\Lambda$CDM. Moreover, the linear growth factor of early dark
energy model is suppressed significantly by the large value of
parameter $C$. When the fraction of dark energy density becomes
large in the EDE model, the present value of $\sigma_8$ will be
lower, $\sigma_8=0.716\pm0.072~(1\,\sigma)$, which is obviously
lower than one obtained in the pure $\Lambda$CDM framework:
$\sigma_8=0.797\pm0.035~(1\,\sigma)$, while the error bar is
enlarged by a factor of two \cite{xiaviel09}. In Fig.~\ref{fig15} we
can clearly see the anti-correlation between $\sigma_8$ and
$\Omega_{\rm EDE,sf}$.

Although the ISW data are directly related to the dark energy
parameters and contain information on the low redshift universe
($z<5$), at present the constraints are too weak to offer much
improvement. In fact, the results do not improve significantly when
adding the ISW data, which is shown in Fig.~\ref{fig18} and Table V.
The constraints on $w_0$ and $C$ improve slightly: $w_0 < -0.708$
and $C < 2.623$ at $95\%$ confidence level which are close to the
pure $\Lambda$CDM model. Meanwhile, the constraints on other
cosmological parameters, such as $\sigma_8$ and $\Omega_{\rm
EDE,sf}$, also tighten a little, but not significantly. In
Fig.~\ref{fig15} we plot the two dimensional constraint on
($\sigma_8$,$\,\Omega_{\rm EDE,sf}$) from different data
combinations. WMAP5+ISW data combination gives clearly tighter
constraints than WMAP5 only.

Finally, we add some low-redshift observational data, such as SNIa
and BAO data. Due to their constraining power, the constraint on
$w_0$ improves significantly: $w_0<-0.909$ at $95\%$ confidence
level. However, the $95\%$ upper limit on $C$ has not been improved:
$C<3.214$, because of the anti-correlation between $w_0$ and $C$
\cite{xiaviel09}. The constraints on other parameters, when
combining all datasets together, become slightly more stringent and
are consistent with the previous work \cite{xiaviel09}, in which a
similar analysis was carried using gamma-ray bursts and
Lyman-$\alpha$ forest observations instead of the ISW effect.

\section{Future Results}
\label{futresult}

From the results presented above, we see that both in $\Lambda$CDM
and EDE model, the ISW data give a little improvement on the
constraints on the cosmological parameter we considered, when
compared to other observations, such as SNIa and BAO. Therefore, it
is worthwhile discussing whether future ISW data could give more
stringent constraints on the cosmological parameters. For this
purpose we have performed a further analysis and we have chosen two
fiducial models in perfect agreement with current data: a pure
$\Lambda$CDM model and an EDE model with parameters taken to be the
best-fit values from the current constraints of ``All'' datasets
combination.


\subsection{$\Lambda$CDM Model}
\label{futerr}

\begin{table}
TABLE VI. Constraints on the $\Lambda$CDM model from the future
measurements. Here we show the standard deviations.
\begin{center}

\begin{tabular}{lccc}

\hline\hline

~Parameter~~&~~CMB~~~&~~CMB+ISW~~&~~All Datasets~~\\
\hline
\multicolumn{4}{c}{$\Lambda$CDM Model}\\

~$b$ & $-$ & $-$ & $0.1943$\\
~$a$ & $-$ & $-$ & $0.0016$\\
~$A_{\rm amp}$ & $-$ & $-$ & $0.2856$\\

~$\Omega_{\rm m}$ & $0.0090$ & $0.0080$ & $0.0029$\\
~$\sigma_8$ & $0.0075$ & $0.0070$ & $0.0054$\\
~$H_0$ & $0.7730$ & $0.7012$ & $0.2327$\\
~${\Omega}_{\rm{EDE,sf}}$&$0.0014$ & $0.0013$ & $0.0005$\\
\hline\hline
\end{tabular}
\end{center}
\end{table}

Firstly, we use the fiducial $\Lambda$CDM mock datasets to constrain
the parameters in the $\Lambda$CDM model, as well as other three
parameters, $b$, $a$ and $A_{\rm amp}$. In Table VI we list the
standard deviations of those parameters from these mock future
measurements. We remind that the mock ISW data sets consist of data
with covariance matrix reduced by a factor nine when compared to the
present ACF and CCF data.

Due to the smaller error bars of the mock data sets, the constraints
on the cosmological parameters from CMB only improve significantly
by a factor of three, when comparing to the current results. When
adding the simulated ISW data, the constraints improve further and
the improvements are larger than those from the current
observations, since the ISW data with smaller error bars are now
more helpful in breaking the degeneracies between the parameters
that keep the CMB angular diameter distance unchanged.

When using all datasets together, we present the constraints on the
cosmological parameters, as well as those of three parameters. The
constraints on the parameters improve significantly. In particular,
due to the more accurate ISW data, the standard deviations of $b$,
$a$ and $A_{\rm amp}$ have been shrunk to $0.19$, $0.15\%$ and
$0.29$. In this case, the significance of ISW signal will be clearly
enhanced. The future CMB measurement and galaxy survey could be very
useful to detect the ISW effect at a much higher significance than
now and also to constrain the bias at high redshift.

\begin{table*}
TABLE VII. Constraints on the early dark energy model from the
future measurements. Here we show the standard deviations. For some
parameters that are only weakly constrained we quote the $95\%$
upper limit.
\begin{center}

\begin{tabular}{l|ccc|ccc}

\hline\hline

~Parameter~~&~~CMB~~~&~~CMB+ISW~~&~~All Datasets~~&~~CMB~~~&~~CMB+ISW~~&~~All Datasets~~\\
\hline
&\multicolumn{3}{|c|}{Fiducial $\Lambda$CDM Model}&\multicolumn{3}{c}{Fiducial EDE Mocker Model}\\

~$w_0$ & $<-0.4952$ & $<-0.6586$ & $<-0.9571$& $<-0.6154$ & $<-0.6455$ & $<-0.9041$\\
~$C$ & $<1.4452$ & $<1.8473$ & $<2.7632$& $<3.3496$ & $<1.9581$ & $<2.4893$\\
~$\Omega_{\rm m}$ & $0.0642$ & $0.0412$ & $0.0031$& $0.0494$ & $0.0406$ & $0.0033$\\
~$\sigma_8$ & $0.0607$ & $0.0425$ & $0.0108$& $0.0512$ & $0.0433$ & $0.0188$\\
~$H_0$ & $5.3335$ & $3.7914$ & $0.3186$& $5.0357$ & $3.8344$ & $0.4733$\\
~$\Omega_{\rm EDE}(z_{\rm lss})$ & $<0.0063$ & $<0.0057$ & $<0.0035$& $<0.0141$ & $<0.0129$ & $<0.0122$\\
~${\Omega}_{\rm{EDE,sf}}$&$0.0055$ & $0.0039$ & $0.0018$&$0.0060$ & $0.0054$ & $0.0043$\\
\hline\hline
\end{tabular}
\end{center}
\end{table*}

\begin{figure}[t]
\begin{center}
\includegraphics[scale=0.41]{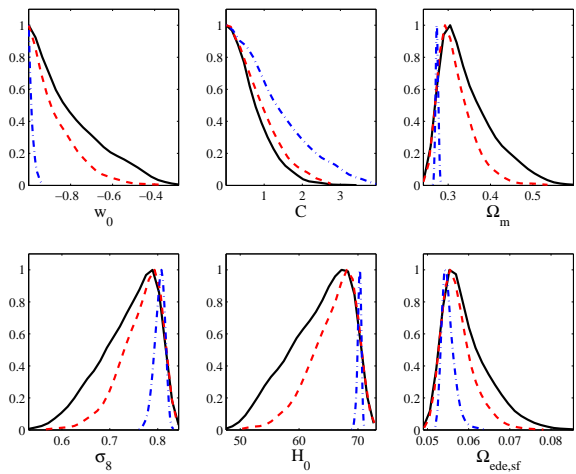}
\caption{One dimensional marginalized likelihood constraints on the
dark energy parameters $w_0$ and $C$, as well as $\Omega_{\rm m}$,
$H_0$, $\sigma_8$ and $\Omega_{\rm EDE,sf}$ from different mock
futuristic data combinations with the fiducial $\Lambda$CDM model:
CMB only (black solid lines), CMB+ISW (red dashed lines) and All
datasets (blue dash-dot lines).\label{fig17}}
\end{center}
\end{figure}

\subsection{EDE Mocker Model}
\label{futede}

\subsubsection{Fiducial $\Lambda$CDM}

Firstly, we choose the pure $\Lambda$CDM as the fiducial model. In
table VII we list the forecasts for some related parameters using
the future measurements.

From Fig.~\ref{fig17} and Table VII, we can find that the CMB data
only cannot constrain the cosmological parameters well, as we
expect. In fact, the standard deviations of parameters are rather
large. Early dark energy models with a non-negligible fraction of
dark energy density cannot be ruled out by the CMB data only,
namely the constraint on the fraction of dark energy density at $z
\sim 1090$ is $\Omega_{\rm EDE}(z_{\rm lss}) < 0.0063$ ($95\%$
C.L.).

Interestingly, when including the future ISW data, the constraints
on the parameters improve significantly, unlike for the current
results. The $95\%$ upper limit of current equation of state of dark
energy $w_0$ is now $w_0 < -0.659$, while $w_0<-0.495$ ($95\%$ C.L.)
obtained from  CMB data only. By contrast, the constraint on $C$ has
not been improved significantly: $C<1.847$ ($95\%$ C.L.), due to the
anti-correlation between $w_0$ and $C$. Meanwhile, all of the
related parameters have been constrained more tightly than ones from
CMB data only  by a factor of two. This simulated ISW data with
smaller error bars is effective in breaking the degeneracies among
the parameters.

Furthermore, when using all datasets, from Fig.~\ref{fig17} and
Table VII we can find that the constraints of many parameters have
been tightened significantly and the degeneracies have been broken
further. The constraint on $w_0$ becomes very tight $w_0<-0.957$ at
$95\%$ confidence level, due to the accurate Supernovae data. And
since the $95\%$ upper limit of $\Omega_{\rm EDE}(z_{\rm lss})$ is
also suppressed apparently, $\Omega_{\rm EDE}(z_{\rm lss}) <
0.0035$, many early dark energy models can be ruled out.

\subsubsection{Fiducial EDE}

\begin{figure}[t]
\begin{center}
\includegraphics[scale=0.41]{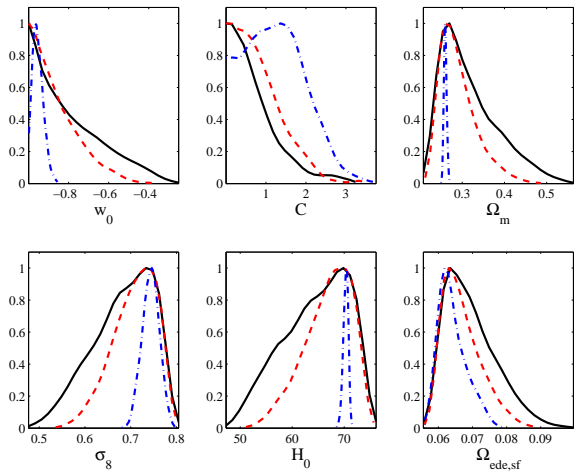}
\caption{One dimensional constraints on the dark energy parameters
$w_0$ and $C$, as well as $\Omega_{\rm m}$, $H_0$, $\sigma_8$ and
$\Omega_{\rm EDE,sf}$ from different mock futuristic data
combinations with the fiducial EDE model: CMB only (black solid
lines), CMB+ISW (red dashed lines) and All datasets (blue dash-dot
lines).\label{fig18}}
\end{center}
\end{figure}

We also choose the EDE fiducial model to determine the cosmological
parameters. The one-dimensional constraints of some related
parameters from different data combinations are plotted in
Fig.~\ref{fig18}. We obtain similar conclusions to the $\Lambda$CDM case.

The CMB data only cannot constrain the parameters very well. When
including the ISW data, the constraints on the parameters improve
significantly. However, even using CMB+ISW data combination, the
best fit values of $w_0$ and $C$ are still close to the $\Lambda$CDM
model. In this case, early dark energy models cannot be
distinguished from the pure $\Lambda$CDM model.

Finally, the ``All'' datasets combination give the most stringent
constraints on the parameters. In this case, the peaks of one
dimensional distributions of $w_0$ and $C$ are moving away from the
$\Lambda$CDM model, $w_0=-1$, $C=0$, see Fig.~\ref{fig18}. The
$95\%$ confidence level are $w_0 < -0.904$ and $C < 2.489$,
respectively. These results imply that the future measurements could
distinguish between the pure $\Lambda$CDM model and
early dark energy models. 

\section{Conclusions and Discussions}
\label{summary}

Most cosmological probes are sensitive to the behaviour of dark
energy at very low redshift; thus, it is of great interest to
investigate those observable which are able to complement the
analysis with constraints on the dark energy abundance at high
redshifts, close to the onset of cosmic acceleration and possibly to
the regime which is probed by other large scale structure
observables like the Lyman-$\alpha$ forest \cite{vieldelya}. In this
paper we exploited the capabilities of the ISW effect in the high
redshift universe, using the cross-correlation signal between the
one million photometrically selected QSOs of the SDSS DR6 catalog
\cite{richards09} and the CMB maps of the WMAP year 5 satellite
\cite{hinshaw09}.  From the SDSS DR6-QSO catalog we extract the QSO
ACF and estimate the bias and stellar contamination by considering
several redshift ranges between $z=0$ and $z=5$. We have given
particular emphasis to our subsamples at $z>1.5$ where an overall
weak evidence for a non-zero ISW signal is found at the
$1.5\,\sigma$ confidence level. The evidence is instead at the
$2.7\,\sigma$ level if we consider the whole sample and this is in
agreement with investigations based on similar samples
\cite{hoetal08,gianna08}.  Our high-redshift sample has a mean
redshift of $z=2$ and $z=2.3$ if we choose the conservative QSO
sample or the speculative one (sources with no ultra-violet excess),
respectively. This is a new regime when compared to other probes
that have been more extensively used for ISW studies such as
galaxies or X-ray observations.

We have focused on modifications to the standard $\Lambda$CDM
cosmology that results in a non-negligible amount of dark energy in
the structure formation era: these models are generically labeled as
Early Dark Energy models and are characterized by an early departure
for matter dominance in the cosmic expansion, with consequences for
background evolution and therefore structure formation. We found
that at present the results are still rather weak to provide
competitive constraints on the parameters describing either the
$\Lambda$CDM or EDE models, even though constraints at the percent
of sub-percent level on the energy density contribution of this
component in the structure formation era can be achieved. Adding
present high redshift ISW data to CMB data from WMAP does not
improve the constraints significantly. However, we also forecast
future performance of QSO data by assuming a measurements of CCF and
ACF data by reducing the error bars by a factor three and combining
these with mock Planck data. In this case the results are
particularly interesting since the improvement to CMB data alone
when adding the ISW information can result in a factor 1.5 on most
of the cosmological parameters. If we further add some Supernovae
luminosity distance moduli like those that can be provided by the
SNAP satellite the constraints can become even tighter and up to a
factor between 3-10 better than the CMB alone.

Here we summarize our main conclusions in more detail:
\begin{itemize}
\item We compute the QSO Auto-Correlation Function (ACF) from SDSS
DR6-QSO and extract bias and catalog efficiency: we measure the
quasar-matter bias and the typical error on the bias is at the level
of $\sigma_{\rm b}=0.8$ when we split in five bins the sample and of
$\sigma_{\rm b}=0.3$ when we consider the whole sample. We present
results for a more conservative selection of QSOs using also
ultra-violet excess flags and a less conservative selection that
consider all the sources of the catalog.
\item EDE models usually result in higher values for the bias than
those of $\Lambda$CDM by $\sim 10\%$, due to the slower growth of
density perturbations in EDE cosmology that requires a higher bias
to match the observed value.
\item We compute CCF values by cross-correlating the QSO number
density with the CMB temperature and found an evidence at
$2.7\,\sigma$ level for the all sample for an ISW effect. At high
redshift this evidence reduces to $1.5\,\sigma$, which interestingly
compares with the prediction of null detection in $\Lambda$CDM. This
non-vanishing signal could be caused by the large bias, stellar
contamination and large number of quasar sample, or an higher dark
energy abundance at high redshifts. However, at present this signal
is still too weak to distinguish between the $\Lambda$CDM and early
dark energy models.
\item The parameter $A_{\rm amp}$ was used to quantify the disagreement
between theoretical predictions and data. We found that this
disagreement vanishes in the high redshift subsample, either for EDE
or $\Lambda$CDM models and is instead at the $\sim 2\,\sigma$ level
for the low-redshift subsamples.
\item We forecasted futuristic large-scale structure measurement of
the ISW effect using the cross-correlation that can be provided by
photometric surveys. Assuming smaller error bars by a factor of
three, the more accurate ISW data is helpful to break the
degeneracies among some parameters and results in an overall
improvement on mock Planck data alone by a factor 1.5. Furthermore,
the standard derivations of those three parameters, $b$, $a$ and
$A_{\rm amp}$, are also shrunk significantly. The future CMB
measurement and galaxy survey could be very useful to detect the ISW
effect at a much higher significance.
\item Further adding Supernovae measurements improve results by a
factor 3-10 due to the better determination of $H_0$ and
$\Omega_{\rm m}$ that can be provided by SNAP observations.
\end{itemize}

The use of this state-of-the-art QSO catalogue can open up a
completely new window on the high-redshift large scale structure of
the universe and allowing for a quantitative use of the ISW effect
in a regime $z>1.5$, which is at present weakly probed by
observations. Either a confirmation of the $\Lambda$CDM model or
possible departures induced by modified gravity or quintessence
could be of fundamental importance and will be addressed by the
surveys of the near future.

\section*{Acknowledgments}

The numerical analysis has been performed on the COSMOS
supercomputer at DAMTP and on Darwin Supercomputer of the University
of Cambridge High Performance Computing Service
(http://www.hpc.cam.ac.uk/), provided by Dell Inc. using Strategic
Research Infrastructure Funding from the Higher Education Funding
Council for England. Some of the results in this paper have been
derived using the HEALPix package \cite{healpix}. We thank G. De
Zotti, T. Giannantonio, A. Myers, C. Morossi, A. Raccanelli, and G.
Richards for useful discussions. We also thank the referee for the
very helpful report. This research has been partially supported by
ASI Contract No. I/016/07/0 COFIS and the INFN-PD51 grant.  MV
thanks the Institute of Astronomy in Cambridge (UK) for hospitality
when part of this work was done.

\end{document}